\documentclass[11pt]{article}
\usepackage[a4paper,margin=1in, centering,ignorehead,nomarginpar]{geometry}
\usepackage[T1]{fontenc}
\usepackage{textcomp}
\usepackage{xcolor}
\usepackage{setspace}
\usepackage{url}
\usepackage{hyperref}
\definecolor{dred}{HTML}{80001F}
\definecolor{dblue}{HTML}{0033BE}
\definecolor{dpurple}{HTML}{57058B}
\hypersetup{%
colorlinks=true,
citecolor=dblue,
linkcolor=dblue,
urlcolor=dblue
}
\usepackage{doi}

\usepackage{array}
\usepackage{natbib}
\bibpunct{(}{)}{;}{a}{,}{;}
\usepackage{graphicx}
\usepackage{bbm}
\usepackage{amsmath,amsthm,amssymb,amsfonts,latexsym}
\usepackage{mathtools}
\usepackage{booktabs}
\usepackage{dcolumn}
\usepackage{icomma}
\usepackage{afterpage}
\usepackage{titling}
\thanksmarkseries{alph}
\usepackage[labelfont={small,sc},textfont={small,it},
            margin=15pt,skip=10pt,position=bottom]{caption}
\usepackage[labelfont={scriptsize,normalfont},textfont={scriptsize,it},
            margin=20pt,skip=5pt,position=bottom,
            labelformat=simple]{subcaption}
\usepackage[running,mathlines]{lineno}
\newcommand*\patchAmsMathEnvironmentForLineno[1]{
  \expandafter\let\csname old#1\expandafter\endcsname\csname #1\endcsname
  \expandafter\let\csname oldend#1\expandafter\endcsname\csname end#1\endcsname
  \renewenvironment{#1}
  {\linenomath\csname old#1\endcsname}
  {\csname oldend#1\endcsname\endlinenomath}}
  \newcommand*\patchBothAmsMathEnvironmentsForLineno[1]{
  \patchAmsMathEnvironmentForLineno{#1}
  \patchAmsMathEnvironmentForLineno{#1*}}
  \AtBeginDocument{
  \patchBothAmsMathEnvironmentsForLineno{equation}
  \patchBothAmsMathEnvironmentsForLineno{align}
 \patchBothAmsMathEnvironmentsForLineno{flalign}
  \patchBothAmsMathEnvironmentsForLineno{alignat}
  \patchBothAmsMathEnvironmentsForLineno{gather}
  \patchBothAmsMathEnvironmentsForLineno{multline}
}

\usepackage{algorithmic}
\usepackage[boxed]{algorithm}

\newtheorem{remark}{Remark}

\newtheorem{proposition}{Proposition}

\numberwithin{condition}{section}
\numberwithin{assumption}{section}
\numberwithin{remark}{section}
\numberwithin{equation}{section}
\numberwithin{lemma}{section}
\numberwithin{definition}{section}
\numberwithin{theorem}{section}
\numberwithin{proposition}{section}
\numberwithin{table}{section}
\numberwithin{figure}{section}
\numberwithin{theorem}{section}
\numberwithin{corollary}{section}
\numberwithin{property}{section}
\numberwithin{algorithm}{section}

\newcommand{\EQ}{\begin{equation}}
\newcommand{\EN}{\end{equation}}
\newcommand{\EQS}{\begin{equation*}}
\newcommand{\ENS}{\end{equation*}}
\newcommand{\ds}{\displaystyle}

\newcommand*\xbar[1]{%
  \hbox{%
    \vbox{%
      \hrule height 0.5pt
      \kern0.5ex%
      \hbox{%
        \kern-0.1em%
        \ensuremath{#1}%
        \kern-0.1em%
      }%
    }%
  }%
}

\def\n1{n}
\def\argmax{\mathop{\rm arg\,max}}

\newsavebox{\savepar}

\numberwithin{equation}{section}
\numberwithin{table}{section}
\numberwithin{figure}{section}

\def\argmax{\mathop{\rm arg\,max}}

\begin{document}
\title{
Optimal control of the decumulation of a retirement portfolio with
variable spending and dynamic asset allocation
}

\author{Peter A. Forsyth\thanks{David R. Cheriton School of Computer Science,
        University of Waterloo, Waterloo ON, Canada N2L 3G1,
        \texttt{paforsyt@uwaterloo.ca}, +1 519 888 4567 ext.\ 34415.}
        \and
        Kenneth R. Vetzal\thanks{School of Accounting and Finance,
        University of Waterloo, Waterloo ON, Canada N2L 3G1,
        \texttt{kvetzal@uwaterloo.ca}, +1 519 888 4567 ext.\ 46518.}
     \and
      Graham Westmacott\thanks{PWL Capital,
        20 Erb Street W., Suite 506,
        Waterloo, ON, Canada N2L 1T2,
        \texttt{gwestmacott@pwlcapital.com}, +1 519 880 0888.}
}

\maketitle

%\linenumbers %Turn on line numbering.

\begin{abstract}
We extend the Annually Recalculated Virtual Annuity (ARVA) spending
rule for retirement savings decumulation \citep{Waring2015} to include
a cap and a floor on withdrawals. With a minimum withdrawal constraint,
the ARVA strategy runs the risk of depleting the investment portfolio.
We determine the dynamic asset allocation strategy which maximizes a
weighted combination of expected total withdrawals (EW) and expected
shortfall (ES), defined as the average of the worst five per cent of
the outcomes of real terminal wealth. We compare the performance of
our dynamic strategy to simpler alternatives which maintain constant
asset allocation weights over time accompanied by either our same
modified ARVA spending rule or withdrawals that are constant over time
in real terms. Tests are carried out using both a parametric model of
historical asset returns as well as bootstrap resampling of historical
data. Consistent with previous literature that has used different
measures of reward and risk than EW and ES, we find that allowing some
variability in withdrawals leads to large improvements in efficiency.
However, unlike the prior literature, we also demonstrate that further
significant enhancements are possible through incorporating a dynamic
asset allocation strategy rather than simply keeping asset allocation
weights constant throughout retirement.

\vspace{5pt}

\textbf{Declarations of interest: } None\\
\textbf{Funding: } Access to Wharton Research Data Services and historical
data from the Center for Research in Security Prices was provided through
an institutional subscription paid for by the University of Waterloo.
Peter Forsyth was also supported by the Natural Sciences
and Engineering Research Council of Canada (NSERC) under grant RGPIN-2017-03760.
\end{abstract}

\newpage
\section{Introduction}
Defined Benefit (DB) pension plans are disappearing, being replaced by
Defined Contribution (DC) plans. According to a recent study by the
Organization for Economic Co-operation and Development (OECD), less
than 50\% of pension assets in 2018 were held in DB plans in over 80\%
of reporting jurisdictions. Moreover, in more than 75\% of reporting
countries the proportion of pension assets in DB plans was lower in 2018
relative to its level a decade earlier \citep{OECD:2019}. Note that the
proportion of assets in DB plans is a lagging indicator of the shift to
DC plans because employees who were historically covered by traditional
DB plans have had more time to amass retirement savings. For example, in
Israel the proportion of pension assets in DB plans dropped from 84\% in
2008 to 56\% in 2018. However, DB plans in that country were closed to
new members in 1995 \citep{OECD:2019}. Almost 25 years later, over half
of pension assets in Israel are still in DB plans.

The shift to DC plans is an inevitable consequence of corporations and
governments being unwilling (or unable) to manage the risks associated
with DB plans. In contrast, in DC plans the management of the financial
assets is left up to individual investors. Given the long-term nature
of retirement savings, this is a challenging task for most people.
Assuming that investors do manage to accumulate healthy balances in
their DC accounts, the situation gets even more complex upon retirement.
Individuals must continue to manage their financial assets, and also
determine a decumulation strategy to withdraw assets and fund spending
with uncertain longevity. While it is often suggested that retirees
should purchase annuities, this rarely happens in practice. For example,
\citet{milevsky-young:2007} report findings from a survey of U.S.
retirees indicating that only 8\% of respondents who were DC plan
members and less than 2\% of all respondents chose to annuitize.
More recently, it has been reported that only around 4\% of retirees
with DC plans at a prominent Canadian insurer opted to annuitize
\citep{carrick:2020}.

The reluctance of retirees to annuitize is sometimes called a puzzle,
since standard life cycle economic models based on utility maximization
suggest that annuitization is optimal \citep{Peijnenburg2016}.
However, the overwhelming aversion to annuitization by retirees
suggests that these economic models are missing something important.
In practice, there are many reasons why retirees do not annuitize.
\citet{MacDonald2013} list dozens of real-world factors including lack
of true inflation protection, loss of control over capital, expensive
pricing, the availability of other sources of guaranteed income such as
government benefits, and paltry payments under some financial market
conditions such as the current low interest rate environment.

Assuming that purchasing an annuity is undesirable, retirees must devise
suitable decumulation strategies. A major component of these plans is
how much money to withdraw over time. Retirees who withdraw fairly
large sums run the risk of outliving their resources, i.e.\ the risk
of ``ruin''. On the other hand those who take out relatively small
amounts may have less enjoyable retirements and leave their heirs with
(unintended) large bequests.

Absent any annuitization, decumulation strategies can generally
be classified as having fixed or variable withdrawals. Within
these categories, several variations have been proposed.
\citet{MacDonald2013} provide a nice summary of the various
possibilities.\footnote{\citet{MacDonald2013} also discuss hybrid
strategies, which combine some level of annuitization with a (fixed or
variable) decumulation scheme. We concentrate on strategies involving
cash flows in the absence of any actual level of annuitization, so
we ignore hybrid strategies in this work.} In a fixed scheme, the
amounts taken out each year are constant, typically in real (i.e.\
inflation-adjusted) terms. This results in a smooth profile of spending
over time, assuming that the retiree remains solvent. In other words,
the risk is effectively due to longevity: the danger is that there will
not be sufficient funds to sustain a very long retirement period with
fixed annual withdrawals. With a variable scheme, the amounts taken out
fluctuate in response to factors such as investment returns. An extreme
example of this would be a fixed percentage withdrawal strategy: the
investor takes out a constant percentage of the portfolio value each
year. In principle, this puts all of the risk onto the spending stream.
It is impossible to run out of funds since something is always left for
the next year. The obvious problem is that the amount withdrawn may fall
below a minimally viable threshold if the retiree lives long enough.
There are many other possibilities for variable schemes which attempt
to strike a balance between the two fundamental risks of spending
fluctuations and longevity, typically through changes in spending in
response to financial market returns.

Perhaps the best known decumulation strategy is the \emph{4\% rule}
due to \citet{Bengen1994}. This fixed scheme states that retirees with
an annually rebalanced portfolio split evenly between bonds and stocks
can withdraw 4\% of their initial wealth each year in real terms.
Backtesting this rule on U.S. data showed that retirees would never have
run out of funds, over any rolling historical 30-year period considered
\citep{Bengen1994}.

Backtesting using rolling historical periods is common in the
practitioner literature. However, in general this approach seriously
underestimates risk. Any two adjacent 30-year periods will have 29
years in common, any two 30-year periods beginning two years apart
will have 28 years in common, etc. Consequently, the overall results
will tend to be highly correlated, and this could be very misleading.
The findings reported by \citet{Bengen1994} address the question of
what the historical experience would have been over a long period for
someone who retired in a particular year and then followed the 4\%
rule. In other words, using rolling historical periods only considers
what \emph{did} happen, giving zero weight to any other plausible
scenario that \emph{might} have happened, and which could occur in
the future. Two alternatives which can give a better sense of the
risk involved are (i) to fit a parametric model to the historical
data and then run a large number of Monte Carlo simulations, and (ii)
to use block bootstrap resampling of the data \citep{politis1994},
which involves randomly drawing (with replacement) shorter periods of
data and chaining them together over the decumulation horizon. We use
both of these approaches below and find that the risk of using the
4\% rule is quite significant.\footnote{There are other reasons to
think that \citet{Bengen1994} understated the risk of the 4\% rule.
One is that data past 1992 was extrapolated using historical averages
for financial market returns each year. For example, the 30-year
performance of the rule given a retirement date of 1976 was assessed
using 16 years of actual market data, followed by 14 years in which the
returns for stocks and bonds and the inflation rate were constant each
year at their long-term average values. This clearly understates the
strategy's risk for cases with several years in retirement after 1992.
A more fundamental issue from today's perspective is the reliability
of the 4\% rule during a lengthy period of very low interest rates.
\citet{finke-pfau-blanchett:2013} considered bond market conditions
early in 2013 and estimated that the failure rate for the 4\% rule
assuming 10 years of below average bond returns and a 50\% stock
allocation was 32\%, strongly suggesting that 4\% is too high a
withdrawal rate. Given that interest rates have continued to trend
downwards more recently, there are solid grounds for pessimism about the
viability of the 4\% rule today.}

As mentioned above, practitioners have proposed several variable schemes
that allow spending to fluctuate in response to portfolio returns. These
strategies typically permit higher initial withdrawal rates compared
to fixed schemes such as the 4\% rule. These enhanced withdrawal rates
can be increased even further following portfolio gains, but need to be
reduced (sometimes severely) after portfolio losses. \citet{Bengen:2001}
considers fixed percentage withdrawals augmented with a floor and
ceiling. The initial withdrawal rate can be increased in line with
investment returns up to a maximum of 25\% higher in real terms than
the first withdrawal, or reduced no further than 10\% below the real
value of the initial withdrawal. \citet{Bengen:2001} concludes that this
strategy permits a safe initial withdrawal rate of about 4.6\%, notably
higher than the fixed 4\% rule. \citet{Guyton-Klinger:2006} explore
the use of a complicated set of heuristic rules governing withdrawals,
portfolio decisions, caps and freezes on inflation adjustments, etc.
They conclude that an initial withdrawal rate of 5.2\%-5.6\% is
sustainable given a portfolio equity allocation of 65\%. As a third
example, \citet{Waring2015} introduce the Annually Recalculated Virtual
Annuity (ARVA) rule, which is based on the idea that the amount taken
out of the portfolio in any given year should be based on the annual
cash flow from a virtual (i.e.\ imaginary) fixed term annuity that could
be purchased using the current value of the portfolio. This strategy is
similar to a fixed percentage withdrawal scheme in that the portfolio
can never be fully depleted, but withdrawals can become unsustainable
small if retirement is sufficiently long and/or portfolio returns are
poor. Alternatively, the ARVA rule will lead to increased withdrawals
following good investment returns.

\citet{Pfau_2015} compares the performance of several spending
strategies by Monte Carlo simulation with parameters calibrated to long
term (1890-2013) annual data for financial market returns and inflation.
\citet{Pfau_2015} begins with a modification of the \citet{Bengen1994}
rule which uses constant inflation-adjusted withdrawals, but with a
spending rate of 2.86\% rather than 4\%. This lower rate of 2.86\%
was estimated on the basis of there being at least a 90\% chance of
1.5\% of the initial amount of real wealth remaining after 30 years
of withdrawals, assuming a 50/50 portfolio allocation between stocks
and bonds. Using the same portfolio allocation and the same 90\%
criterion for other strategies permitted higher initial spending rates.
For example, the initial spending rate for \citet{Bengen:2001}'s
fixed percentage scheme with a floor of 85\% of the real value of the
first year's withdrawal and a corresponding ceiling of 120\% resulted
in a sustainable initial spending rate of 3.31\%. As additional
examples, \citet{Pfau_2015}'s implementations of the ARVA approach
\citep{Waring2015} and the \citet{Guyton-Klinger:2006} rules produced
sustainable initial spending rates of 4.34\% and 4.82\% respectively.

An important issue that has not been investigated much in the
practitioner literature on decumulation is the effect of a more
sophisticated approach to asset allocation, beyond simply rebalancing to
a constant weighting of bonds and stocks. \citet{Yamada_2017} explore
the performance of rebalancing to maintain a constant level of a
(time-varying) equity market risk measure using several withdrawal rules
and report that sustainable spending is significantly improved. However,
this leaves open the question of the impact of using an asset allocation
strategy that is optimized to achieve a well-defined financial
objective. Implementing such an approach necessitates specifying a
suitable objective function and solving the resulting optimization
problem, which in turn requires more technically sophisticated methods.

Along these lines, \citet{dang_2016} suggest using a multi-period mean
variance objective function to examine the effect of different (fixed)
withdrawal rates coupled with an adaptive portfolio allocation strategy.
The objective function is posed in terms of the mean and variance of the
final wealth at time $T$. \citet{dang_2016} assume that most 65-year
olds can expect to live for 20 years with high probability, and thus set
a wealth target of one-half of the initial wealth at $T=20$ years (after
retirement). The idea is that retirees can decide how to hedge longevity
risk at age $85$, expecting to have spent one-half of their original
wealth up to then.

\citet{Irlam:2014} uses dynamic programming methods to determine
asset allocation, given an objective of maximizing the number of
years of solvency divided by the number of years lived. This is the
only study we are aware of in the practitioner literature for which
the asset allocation depends on a specified financial objective.
\citet{Irlam:2014} concludes that asset allocation rules that depend
only on time such as ``age in bonds'' or various target-date fund
glide paths require a higher amount of investment in order to obtain
the same withdrawal rates in retirement, as compared to his approach
where the asset allocation is time and state-dependent. However,
\citet{Irlam:2014} only considers a fixed annual withdrawal amount in
retirement.

In this work we further explore the effect of a variable spending rule
in combination with an asset allocation strategy tailored to optimizing
a financial objective. In particular, we use an ARVA spending rule
augmented by constraints on minimum and maximum
annual withdrawals. The minimum withdrawal constraint means that
there is risk of depleting the portfolio entirely prior to the end
of the investment horizon. We use the Expected Shortfall (ES) of the
terminal portfolio value as a measure of risk. The ES at level $x\%$
is the mean of the worst $x\%$ of outcomes, and is thus a measure of
tail risk. As a measure of reward, we use total Expected Withdrawals
(EW). Based on a parametric model calibrated to historical market data,
we determine the portfolio allocation strategy that optimizes the
multi-objective Expected Withdrawals-Expected Shortfall (EW-ES)
objective function.\footnote{\citet{Forsyth_2019_x} use the same measure
of reward, but minimize the downside variability of withdrawals for an
ARVA type spending rule, i.e.\ the risk measure is downward withdrawal
variability. There are some other noteworthy differences between this
work and that of \citet{Forsyth_2019_x}. First, we impose upper and
lower bounds on annual withdrawals. Second, the assumed underlying
financial model is more complex here, as it incorporates stochastic bond
market returns.}

A similar measure of risk and reward for DC plan decumulation is used
in \citet{Forsyth_2020_xx}.  However, \citet{Forsyth_2020_xx} uses the withdrawal
amount as a control, rather than an ARVA spending rule.  In this case,
\citet{Forsyth_2020_xx} shows that the withdrawal control is essentially
a bang-bang type control, with minimum withdrawals during the earlier
years of retirement.  Use of the ARVA spending rule (with constraints)
provides more control over the timing of withdrawals.

We verify the robustness of this strategy through tests using bootstrap
resampling of historical return data. Our tests show that the ARVA
spending rule coupled with an optimal allocation strategy is always more
efficient than a constant withdrawal, constant weight strategy. In fact,
our optimal dynamic ARVA strategy outperforms this alternative even
when the \emph{minimum} withdrawal under ARVA is equal to the constant
withdrawal with constant weights. This verifies that allowing some
variability in withdrawals sharply reduces the risk of depleted savings,
consistent with \citet{Pfau_2015} and \citet{Yamada_2017}. In addition,
we demonstrate that solving an optimal stochastic control problem to
specify the asset allocation can provide further significant benefits
beyond those obtained by permitting withdrawal variability alone.

\section{ARVA Spending Rule}\label{ARVA_section}
Consider the following spending rule.   Each year, a
virtual (hypothetical) fixed term annuity is constructed, 
based on the current portfolio value,
the number of remaining years of required cash flows, and a
real (inflation adjusted) interest rate. The investor then 
withdraws an  amount based on the hypothetical payment 
of this virtual annuity.
Clearly, the annual payments will be variable,
since the virtual annuity is
recalculated each year, and is a function of the current portfolio value.  The portfolio is liquidated
at the end of the investment horizon.  A surplus will
be returned to the investor (or the investor's estate).
Any shortfall must be settled at this time as well.

We are now faced with the choice of determining a timespan for the
virtual fixed term annuity. Rather than specifying a maximum possible
lifespan (which would be overly conservative), we assume that retirees
are in the top 20\% of the population in terms of conditional expected
longevity \citep{West2017}. Consider a retiree who is $x$ years old at
$t=0$. Assuming that the $x+t$ year old retiree is alive at time $t$,
let $T^*_x(t)$ be the time at which 80\% of the cohort of $x+t$ year
olds are expected to have passed away, conditional on all members of
the cohort being alive at time $t$. At time $t$, the fixed term of the
virtual annuity is then $T^*_x(t)-t$. This mortality assumption has
the effect of providing increased spending during the early years of
retirement. By varying the fraction of the cohort assumed to have passed
away, we can increase/decrease spending in early retirement years at the
cost of decreased/increased spending in later years. Note that our ARVA
withdrawal amount is not generally the same as would be obtained from a
currently purchased life annuity.

Given the real interest rate $r$, the present value of an annuity which
pays continuously at a rate of one unit per year for $T^*_x(t)-t$
years is denoted by the annuity factor
\begin{equation}
a(t) = \frac{1 - \exp[-r(T^*_x(t)-t)]}{r} ~.
\end{equation}
It follows that $W(t)/a(t)$ is the continuous real annuity payment for $(T^*_x(t) - t)$ years,
which can be purchased with wealth $W(t)$ at time $t$.
We make the assumption that withdrawals occur at
discrete times in
\begin{equation}
\mathcal{T} \equiv \{t_0 = 0 < t_1 < \dots < t_M = T\},
\label{eq:intervention_time}
\end{equation}
where $t_0$ denotes the time that the $x$ year old retiree begins to
withdraw money from the DC plan. We assume the times in
$\mathcal{T}$ are equally spaced with $t_i - t_{i-1} = \Delta t = T/M$,
$i = 1, \dots, M$. We let $\Delta t = $ one year. 
We determine the cash withdrawal at time $t_i$ by converting
the continuous payment above into a lump sum received in advance of the
interval $[t_i, t_{i+1}]$. This lump sum withdrawal at $t_i$ is $W(t_i)
A(t_i)$, where
\begin{equation}
A(t_i) = \int_{t_i}^{t_{i+1}} 
           \frac{~e^{-r(t^{\prime}-t_i)}}{a(t^{\prime})}\, dt^{\prime}.
\label{annuity_factor_1}
\end{equation}
In this work, we will compute equation (\ref{annuity_factor_1})
based on
the CPM 2014 mortality tables
(male) from the Canadian Institute of
Actuaries\footnote{\url{www.cia-ica.ca/docs/default-source/2014/214013e.
pdf}} to compute $T^*_x(t)$ with $x=65$. Further discussion of the ARVA
spending rule can be found in \citet{Forsyth_2019_x}.

\section{Investment Market}
We assume that the investment portfolio consists of
two index funds.  These funds include a stock market index fund
and a constant maturity bond index fund.
Let the investment horizon be $T$, and  $S_t$ and $B_t$ respectively denote the real
(inflation adjusted) amounts invested in the stock index and the bond
index. These amounts can change due to (i) changes in
the real unit prices and (ii) the investor's asset allocation
strategy.
In the absence of the application of an  investor's control,
all changes in $S_t$ and $B_t$
result from changes in asset prices. 

We model the stock index (in the absence of an applied control) 
as following a jump diffusion process. Let
$S_{t^-} = S(t-\epsilon), \epsilon \rightarrow 0^+$, i.e.\ $t^-$ is
the instant of time before $t$, and let $\xi^s$ be a random jump
multiplier. When a jump occurs, $S_t = \xi^s S_{t^-}$. Use of
jump processes allows for 
modelling of fat-tailed (non-normal) asset
returns.\footnote{Appendix~\ref{calibration_section} documents evidence
of leptokurtic behavior for both of the indexes that we use in our
tests.} We assume that $\log(\xi^s)$ follows a double exponential
distribution \citep{Kou2004}. The probability of an upward
jump is $p_{\text{\emph{u}}}^s$, with $1-p_{\text{\emph{u}}}^s$ being the
probability  of a downward jump. The density function for $y = \log (\xi^s)$
is
\begin{equation}
f^s(y) = p_{u}^s \eta_1^s e^{-\eta_1^s y} {\bf{1}}_{y \geq 0} +
       (1-p_{u}^s) \eta_2^s e^{\eta_2^s y} {\bf{1}}_{y < 0}.
\label{eq:dist_stock}
\end{equation}
Define
\begin{equation}
\kappa^s_\xi = E[\xi^s -1]
         = \frac{p_{\text{\emph{{u}}}}^s \eta_1^s}{\eta_1^s - 1} + 
           \frac{(1-p_{\text{\emph{{u}}})}^s \eta_2^s}{\eta_2^s + 1} - 1.
\end{equation}
Without an applied control,
\begin{equation}
\frac{dS_t}{S_{t^-}} = \left(\mu^s -\lambda_\xi^s \kappa_{\xi}^s \right) \, dt
   + \sigma^s \, dZ^s + d\left(\ds \sum_{i=1}^{\pi_t^s}(\xi_i^s - 1)\right),
\label{jump_process_stock}
\end{equation}
where $\mu^s$ is the (uncompensated) drift rate, $\sigma^s$ is the
diffusive volatility, $Z^s$ is a Brownian motion, $\pi_t^s$ is
a Poisson process with intensity parameter $\lambda_\xi^s$, and
$\xi_i^s$ are i.i.d.\ positive random variables having distribution
(\ref{eq:dist_stock}). Moreover, $\xi_i^s$, $\pi_t^s$, and $Z^s$ are
assumed to all be mutually independent.

As in \citet{mitchell_2014} and \citet{Lin_2015}, we use a common
practitioner approach and model the returns of the constant maturity
bond index (absent an applied control) as a stochastic process. This
approach has the advantage that estimating model parameters from market
data is quite straightforward, without the need to devise a parametric
process for real interest rates. As in \citet{mitchell_2014}, we assume
that the constant maturity bond index follows a jump diffusion process.
In particular, $B_{t^-} = B(t-\epsilon), \epsilon \rightarrow 0^+$. In
the absence of control, $B_t$ evolves as
\begin{equation}
\frac{dB_t}{B_{t^-}} = \left(\mu^b -\lambda_\xi^b \kappa_{\xi}^b  
  + \mu_c^b {\bf{1}}_{\{B_{t^-} < 0\}}  \right) \, dt + 
  \sigma^b \, dZ^b +  d\left( \ds \sum_{i=1}^{\pi_t^b}(\xi_i^b -1)\right),
\label{jump_process_bond}
\end{equation}
where the terms in equation (\ref{jump_process_bond}) are defined
analogously to equation (\ref{jump_process_stock}). In particular,
$\pi_t^b$ is a Poisson process with positive intensity parameter
$\lambda_\xi^b$, and $\xi_i^b$ has distribution
\begin{equation}
f^b(y=\log \xi^b) = p_{u}^b \eta_1^b e^{-\eta_1^b y}{\bf{1}}_{y \geq 0} +
  (1-p_{u}^b) \eta_2^b e^{\eta_2^b y}{\bf{1}}_{y < 0},
\label{eq:dist_bond}
\end{equation}
and $\kappa_{\xi}^b = E[ \xi^b -1 ]$. $\xi_i^b$, $\pi_t^b$, and
$Z^b$ are assumed to all be mutually independent. The term $\mu_c^b
{\bf{1}}_{\{B_{t^-} < 0\}}$ in equation (\ref{jump_process_bond})
represents an additional cost of borrowing ($B_t < 0$), i.e.\ a spread
between borrowing and lending rates. We assume that the diffusive
components of $S_t$ and $B_t$ are correlated, i.e.\ $dZ^s \cdot
dZ^b = \rho_{sb} ~dt$. However, the jump process terms for these
two indexes are assumed to be mutually independent.\footnote{See
\citet{Forsyth_2020_IME} for a discussion of the evidence for stock and
bond price jump independence.}

It is possible to include more complex stock and bond processes,
such as stochastic volatility for example. However, \citet{Ma2015}
have shown that including stochastic volatility effects does not
have a significant effect on the results for long term investors. In
order to verify the robustness of the strategies, we will determine
the optimal controls using the parametric model based on equations
(\ref{jump_process_stock}) and (\ref{jump_process_bond}). We then test
these controls on bootstrapped resampled historical data. This is quite
a strict test, since the bootstrapped resampling algorithm makes no
assumptions about the underlying bond and stock stochastic processes.

We define the investor's total wealth at time $t$ as $W_t \equiv S_t +
B_t$. We impose the constraints that (assuming solvency) shorting stock
and using leverage (i.e.\ borrowing) are not allowed. Insolvency can
arise from withdrawals. If this happens, the portfolio is liquidated
and debt accumulates at the borrowing rate. The borrowing rate is taken
to be the return on the constant maturity bond index plus a spread
$\mu_c^b$.

\section{Notational Conventions}
\label{adaptive_section}
For ease of explanation,
we will occasionally use the notation
$S_t \equiv S(t), B_t \equiv B(t)$ and $W_t \equiv W(t)$. Earlier in
equation (\ref{eq:intervention_time}) we specified a set of times
$\mathcal{T}$ for which withdrawals are permitted. We now expand the
scope of $\mathcal{T}$ so that portfolio rebalances are also allowed at
those times, i.e.\ $\mathcal{T}$ is the set of withdrawal/rebalancing
times. More specifically, let the inception time of the investment
be $t_0 = 0$. At each withdrawal/rebalancing time $t_i$, $i = 0, 1,
\ldots, M-1$, the investor (i)~withdraws an amount of cash $q_i$ from
the portfolio, and then~(ii) rebalances the portfolio. At $t_M = T$, the
portfolio is liquidated and the final cash flow $q_M$ occurs.

Given a time dependent function $f(t)$, we use the shorthand notation
$f(t_i^+) \equiv \displaystyle \lim_{\epsilon \rightarrow 0^+} f(t_i +
\epsilon)$ and $f(t_i^-) \equiv \displaystyle \lim_{\epsilon \rightarrow
0^+} f(t_i - \epsilon)$. We assume that no taxes are triggered by
rebalancing. This would normally be the case in a tax-advantaged DC
savings account. Since we assume yearly application of the controls
(rebalancing), we expect transaction costs to be small and hence they
can be safely ignored.\footnote{It is possible to include transaction
costs, but this will increase computational cost \citep{Van2018}.} With
no taxes or transaction costs, it follows that $W(t_i^+) = W(t_i^-) -
q_i$.

The multi-dimensional controlled underlying process is denoted
by $X(t) = \left(S\left(t\right), B\left(t\right)\right)$, with
$t\in\left[0,T\right]$. The realized state of the system is $x = (s,b)$.
Let the rebalancing control $p_i(\cdot)$ be the fraction invested in the
stock index at rebalancing date $t_i$, i.e.\
\begin{equation}
p_i \left(X(t_i^-)\right) = p \left(X(t_i^-),t_i \right) = 
  \frac{S(t_i^+)}{S(t_i^+)+B(t_i^+)}.
\end{equation}
The controls depend on the state of the investment
portfolio before the rebalancing occurs, i.e.\ $p_i(\cdot) =
p\left(X(t_i^-),t_i\right) = p\left(X_i^-, t_i \right)$, $t_i
\in \mathcal{T}$.
We search for the optimal strategies amongst all controls with
constant wealth  after cash withdrawal,
\begin{align}
  p_i(\cdot) &= p(W(t_i^+), t_i) \nonumber \\
  W(t_i^+) &= S(t_i^-) + B(t_i^-) - q_i \nonumber \\
  S(t_i^+) &= S_i^+ = p_i(W_i^+) \, W_i^+ \nonumber \\
  B(t_i^+) &= B_i^+ = (1 -p_i(W_i^+))\, W_i^+.  \label{p_def_2}
\end{align}
We assume that rebalancing occurs instantaneously, with the
implication that
the probability of a jump occurring in either index is zero
during the rebalancing period $(t_i^-, t_i^+)$.

Let $\mathcal{Z}$ represent the set of admissible values of the control
$p_i(\cdot)$. An admissible control $\mathcal{P} \in \mathcal{A}$,
where $\mathcal{A}$ is the admissible control set, can be written as
$\mathcal{P} = \{p_i(\cdot) \in \mathcal{Z}~:~ i=0, \ldots, M-1 \}$.
We impose no-shorting and
no-leverage constraints by specifying
\begin{equation}
\mathcal{Z} = [0,1].
\label{no_short_no_leverage} 
\end{equation}
We also apply the constraint that if $W(t_i^+)
< 0$,  the stock index holding is liquidated, 
\begin{equation}
p(W(t_i^+), t_i) = 0 \text{ if } W(t_i^+) < 0,
\label{insolvent_eqn}
\end{equation}
and no further stock purchases are permitted,
with the result that debt accumulates at the
bond return plus a spread.
In addition, we define $\mathcal{P}_n \equiv \mathcal{P}_{t_n} \subset
\mathcal{P}$ as the tail of the set of controls in $[t_n, t_{n+1},
\ldots, t_{M-1}]$, i.e.\ $\mathcal{P}_n =\{p_n(\cdot), \ldots,
p_{M-1}(\cdot)\}$.

\section{Risk and Reward Measures}
Initially, we  describe our measure of risk. Suppose $g(W_T)$ is the
probability density function of terminal wealth $W_T$ at $t=T$, and let
\begin{equation}
\int_{-\infty}^{W^*_{\alpha}}  g(W_T) \, dW_T = \alpha,
\label{CVAR_def_a}
\end{equation}
so that \emph{Prob}$[W_T > W^*_{\alpha}] = 1 - \alpha$. We can interpret
$W^*_{\alpha}$ as the Value at Risk (VAR) at level $\alpha$. The
Expected Shortfall (ES) at level $\alpha$ is then
\begin{equation}
\text{ES}_{\alpha} = \frac{\int_{-\infty}^{W^*_{\alpha}} W_T \, g(W_T) \, dW_T}
                          {\alpha},
\label{ES_def_1}
\end{equation}
which is the mean of the worst $\alpha$ fraction of outcomes. Usually,
$\alpha \in \{ .01, .05 \}$. We emphasize that the definition of ES in equation
\eqref{ES_def_1} uses the probability density of the final wealth
distribution, not the density of \emph{loss}. This has
the implication that 
a larger value of ES is desirable (the worst case average portfolio
value at $T$).\footnote{The negative of ES is often called
Conditional Value at Risk (CVAR), which has been used as a risk measure
in several prior asset allocation studies \citep[e.g.\@][]{gao-xiong-li:2016,
cui-gao-shi-zhu:2019,forsyth_2019_c}.}

Define $X_0^+ = X(t_0^+), X_0^- = X(t_0^-)$. Given an expectation under
control $\mathcal{P}$, $E_{\mathcal{P}}[\cdot]$, \citet{Uryasev_2000}
show that $\text{ES}_{\alpha}$ can be alternatively written as
\begin{equation}
{\text{ES}}_{\alpha}( X_0^-, t_0^-) = 
  \sup_{W^*}E_{\mathcal{P}_0}^{X_0^+, t_0^+}
  \biggl[W^* + \frac{1}{\alpha} \min(  W_T - W^* , 0 )\biggr].
\label{ES_def}
\end{equation}
The notation ${\text{ES}}_{\alpha}(
X_0^-, t_0^-)$ indicates that ${\text{ES}}_{\alpha}$ is as seen at
$(X_0^-, t_0^-)$. This definition is then  the \emph{pre-commitment}
ES. A strategy based on optimizing the pre-commitment ES at
time zero is \emph{time inconsistent}, since the investor may have
an incentive to deviate from the strategy at $t>0$. Thus,
some authors have described pre-commitment strategies as being
\emph{non-implementable}. However, this is really a matter of
interpretation: we consider the pre-commitment strategy as
a useful technique to compute an appropriate value of 
$W^*$ in equation (\ref{ES_def}).  In fact, the strategy
which fixes $W^*$ $\forall t >0$, is the
\emph{induced} time consistent strategy \citep{Strub_2019_a},
and is consequently implementable.
We delay further discussion of this point to Section~\ref{obj_fun}.

Our measure of reward is expected total withdrawals (EW), defined as
\begin{equation}
\text{EW}(X_0^-,t_0^-) = E_{\mathcal{P}_0}^{X_0^+, t_0^+} 
  \biggl[\,\sum_{i=0}^{i=M} q_i \, \biggr].  \label{EW_def}
\end{equation}
Note that we do not discount withdrawals, with either a market-based
measure of the appropriate risk-adjusted discount rate or with a
subjective discount rate. This reflects a desire to avoid basing our
strategy on parameters that are difficult to estimate. Since the
portfolio weights will depend on realized investment returns and
withdrawals over time, it is problematic to estimate the appropriate
risk-adjusted discount rate. Moreover, it is likely to be difficult
to determine a subjective discount rate, which could easily vary
across investors and/or over time. However, we observe that the
economic effect of discounting the withdrawals would be to make earlier
withdrawals more desirable. We have already incorporated a similar
effect through the mortality boost to the spending rule discussed in
Section~\ref{ARVA_section} above.

\section{Objective Function}
\label{obj_fun}
Our overall approach involves a statistical tradeoff between reward and
risk, similar to mean-variance portfolio analysis but with different
measures of reward and risk. The main alternative would be a standard
life cycle approach, where we would maximize a specified utility
function. This would raise concerns related to estimating parameters
such as risk aversion or elasticities of intertemporal substitution,
similar to the subjective discount rate discussed in the preceding
paragraph. However, this would pose more of a problem since the
appropriate form of the utility function itself is open to question. The
most popular specification in the literature is power utility, which
implies constant relative risk aversion. However, a recent empirical
study by \citet{Meeuwis:2020} of the portfolio holdings and income
of millions of US retirement investors indicates that such a model
is mis-specified: actual investors exhibit decreasing (not constant)
relative risk aversion. More generally, the standard life cycle approach
in principle requires knowledge of the investor's total wealth including
wealth due to human capital, illiquid assets such as a home, etc.,
not just a retirement savings portfolio. Although the standard life
cycle approach offers some insightful theoretical implications, it is
difficult to use in practice because the information required is often
either not available or measured very imprecisely. We can also point out
that the empirical validity of the standard life cycle approach has been
questioned on behavioral grounds \citep{thaler:1990}. Accordingly, we
avoid standard life cycle modelling based on utility functions. We also
avoid extending the standard life cycle approach to more complicated
preference specifications which may fit the data better \citep[see,
e.g.\@][and references therein]{Meeuwis:2020}. Instead, we take the
relatively simpler approach of optimizing the reward-risk tradeoff.

Expected withdrawals (EW) and expected shortfall (ES) are conflicting
measures, so we use a scalarization technique to find the Pareto
points for this multi-objective optimization problem. Informally, for
a given scalarization parameter $\kappa > 0$, we seek the control
$\mathcal{P}_0$ that maximizes
\begin{equation}
\text{EW}(X_0^-, t_0^-)  + \kappa \text{ES}_{\alpha}(X_0^-, t_0^-).
\label{eq:informal}
\end{equation}
More precisely, we define the pre-commitment EW-ES problem in terms of
the value function
\begin{equation}
J\left(s,b,t_0^-\right) = 
  \sup_{\mathcal{P}_0\in\mathcal{A}} \sup_{W^*}
  \Biggl\{E_{\mathcal{P}_0}^{X_0^+,t_0^+}
  \Biggl[\sum_{i=0}^M q_i + 
  \kappa\left(W^* + \frac{\min(W_T-W^*,0)}{\alpha}\right) 
  \bigg\vert X(t_0^-)=(s,b)\Biggr] \Biggr\} \label{PCEE_a}
\end{equation}
and the constraints
\begin{align}
&(S_t, B_t) \text{ follow processes \eqref{jump_process_stock} and 
  \eqref{jump_process_bond}}; \quad t \notin \mathcal{T} \nonumber \\
&W_{\ell}^+ = S_{\ell}^- + B_{\ell}^- = q_{\ell}; 
             \quad X_{\ell}^+ = \left(S_{\ell}^+,B_{\ell}^+\right) \nonumber \\
&S_{\ell}^+ = p_{\ell}(\cdot)W_{\ell}^+;
              \quad B_{\ell}^+ = \left(1 - p_{\ell}(\cdot)\right)W_{\ell}^+
              \nonumber \\
&p_{\ell}(\cdot) \in \mathcal{Z} = [0,1] \text{ if } W_{\ell}^+ > 0; \quad
p_{\ell}(\cdot) = 0 \text{ if } W_{\ell}^+ \leq 0 \nonumber \\
&\ell = 0, \dots, M-1; \quad t_{\ell} \in \mathcal{T}.
\label{PCEE_b}
\end{align}
By reversing the order of  the $\sup \sup$ in equation (\ref{PCEE_a}),
the value function can be written as
\begin{equation}
J\left(s,b,t_0^-\right) = \sup_{W^*}\sup_{\mathcal{P}_0 \in\mathcal{A}}
  \Biggl\{E_{\mathcal{P}_0}^{X_0^+,t_0^+}
  \Biggl[\sum_{i=0}^{i=M} q_i +
  \kappa \left(W^* + \frac{\min (W_T -W^*, 0)}{\alpha} \right)
  \bigg\vert X(t_0^-) = (s,b)\Biggr]\Biggr\}.
\label{pcee_inter}
\end{equation}
Denote the investor's initial wealth at $t_0$ by $W_0^- = S_0^- +
B_0^-$. Observe that the inner supremum in equation (\ref{pcee_inter})
is a continuous function of $W^*$. Then, assuming that the domain of
$W^*$ is compact, we define
\begin{multline}
\mathcal{W}^*(0,W_0^-) = \argmax_{W^*}
  \Biggl\{\sup_{\mathcal{P}_0 \in A}
  \Biggl\{E_{\mathcal{P}_0}^{X_0^+,t_0^+}
  \left[\sum_{i=0}^{i=M} q_i + 
  \kappa\left(W^* + \frac{\min(W_T-W^*,0)}{\alpha}\right)
  \right. \Biggr. \Biggr. \\
  \left. \Biggl. \Biggl.
  \bigg\vert X(t_0^-)=(0,W_0^-)
  \right]
  \Biggr\}
  \Biggr\}.
\label{pcee_argmax}
\end{multline}

Regarding $\mathcal{W}^*(0,W_0^-)$ as fixed $\forall t >0$, the following
proposition follows immediately:
\begin{proposition}[Pre-commitment strategy equivalence to a time
consistent policy for an alternative objective function]
\label{equiv_thm}
The pre-commitment EW-ES strategy $\mathcal{P}^*$ determined by
solving $J(0,W_0,t_0^-)$ with $\mathcal{W}^*(0,W_0^-)$ from equation
(\ref{pcee_argmax}) is the time consistent strategy for an equivalent
problem with fixed $\mathcal{W}^*(0,W_0^-)$ and value function
$\tilde{J}(s,b,t)$ defined by
\begin{equation}
\tilde{J}(s,b,t_n^-) = \sup_{\mathcal{P}_n \in \mathcal{A}}
  \Biggl\{E_{\mathcal{P}_n}^{X_n^+,t_n^+}
  \Biggl[\sum_{i=n}^{i=M}q_i + \frac{\kappa\min(W_T-\mathcal{W}^*(0,W_0^-),0)}
                                    {\alpha}
  \bigg\vert X(t_n^-)=(s,b)
  \Biggr]
  \Biggr\}.
  \label{timec_equiv}
\end{equation}
\end{proposition}

\begin{remark}[EW-ES induced time consistent strategy: an implementable control]
In the following, we consider the actual strategy followed by the
investor for any $t>0$ as given by the induced time consistent
strategy\footnote{See \citet{Strub_2019_a} for a discussion of
induced time consistent strategies.} that solves problem~(\ref{timec_equiv}) with the fixed value
of $\mathcal{W}^*(0,W_0^-)$ from equation (\ref{pcee_argmax}). This strategy is identical to the EW-ES
strategy at time zero. Hence, we refer to this strategy as the
EW-ES strategy. It is understood that this refers to the strategy
that solves the time consistent equivalent problem~(\ref{timec_equiv})
for any $t > 0$.
Consequently, this strategy is implementable \citep{forsyth_2019_c} (the investor has no incentive to
deviate from this control for $t >0$ ).
\end{remark}

\section{Solution Method}
\label{algo_section}
To solve the pre-commitment EW-ES problem~(\ref{PCEE_a}), we start by
interchanging the $\sup \sup$ to arrive at equation~(\ref{pcee_inter}).
We expand the state space to $\hat{X} = (s,b,W^*)$, and define the
auxiliary value function
\begin{equation}
V(s,b,W^*,t_n^-) = 
  \sup_{\mathcal{P}_n\in\mathcal{A}} 
  \Biggl\{E_{\mathcal{P}_n}^{\hat{X}_n^+,t_n^+}
  \Biggl[\sum_{i=n}^M q_i + 
  \kappa\left(W^* + \frac{\min(W_T-W^*,0)}{\alpha}\right) 
  \bigg\vert \hat{X}(t_n^-)=(s,b,W^*)\Biggr]\Biggr\}
\label{expanded_1} 
\end{equation}
and slightly revised constraints
\begin{align}
&(S_t, B_t) \text{ follow processes \eqref{jump_process_stock} and 
  \eqref{jump_process_bond}}; \quad t \notin \mathcal{T} \nonumber \\
&W_{\ell} = S_{\ell}^- + B_{\ell}^- = q_{\ell}; 
            \quad \hat{X}_{\ell}^+ = \left(S_{\ell}^+,B_{\ell}^+,W^*\right)
            \nonumber \\
&S_{\ell}^+ = p_{\ell}(\cdot)W_{\ell}^+;
              \quad B_{\ell}^+ = \left(1 - p_{\ell}(\cdot)\right)W_{\ell}^+
            \nonumber \\
&p_{\ell}(\cdot) \in \mathcal{Z} = [0,1] \text{ if } W_{\ell}^+ > 0; \quad
p_{\ell}(\cdot) = 0 \text{ if } W_{\ell}^+ \leq 0 \nonumber \\
&\ell = 0, \dots, M-1; \quad t_{\ell} \in \mathcal{T}.
\label{expanded_2}
\end{align}
We can solve
auxiliary problem (\ref{expanded_1}) using dynamic programming. The optimal control
$p_n(w,W^*)$ at time $t_n$ is determined from
\begin{equation}
p_n(w, W^*) = 
\begin{cases}
\displaystyle
\argmax_{p^{\prime} \in \mathcal{Z}} V(wp^{\prime},w(1-p^{\prime}),W^*,t_n^+)
  &\text{ if } w > 0 \\
0 &\text{ if } w \leq 0
\end{cases}.
\label{expanded_3}
\end{equation}
Following the dynamic programming algorithm, we move the solution backwards
across  across time $t_n$ via
\begin{equation}
V(s,b,W^*,t_n^-) = V(w^+ p_n(w^+,W^*), w^+\left(1 - p_n(w^+,W^*)\right), 
  W^*, t_n^+) + q_n(w^-,W^*)~,
\label{expanded_4}
\end{equation}
where $w^- = s + b $, and $w^+ = w^- - q_n$.   $q_n(w^-,W^*)$ is based on our ARVA spending
rule (see Section \ref{investment_section} for a precise specification).  
Note that the spending rule will be a function of wealth before withdrawal.
At $t=T$, we have
\begin{equation}
V(s,b,W^*,T^+) = \kappa \biggl(W^* + \frac{\min(s+b-W^*,0)}{\alpha}\biggr).
\label{expanded_5}
\end{equation}
For times $t \in(t_{n-1},t_n)$, there are no cash flows or
controls applied. Recall that all quantities are real, and that
there is no discounting. The iterated expectation property
combined with It\^{o}'s Lemma for jump processes in equations
(\ref{jump_process_stock}-\ref{jump_process_bond}) then gives
\begin{align}
V_t &+ \frac{(\sigma^s)^2 s^2}{2}V_{ss} +
       (\mu^s - \lambda_{\xi}^s \kappa_{\xi}^s)sV_s
       + \lambda_{\xi}^s \int_{-\infty}^{+\infty}V(e^ys,b,t)f^s(y)\,dy 
       \nonumber \\ 
    &+ \frac{(\sigma^b)^2 b^2}{2}V_{bb} + 
       (\mu^b - \lambda_{\xi}^b \kappa_{\xi}^b)b V_b
       + \lambda_{\xi}^b \int_{-\infty}^{+\infty}V(s,e^yb,t)f^b(y)\,dy 
       \nonumber \\
    &- (\lambda_{\xi}^s + \lambda_{\xi}^b)V 
       + \rho_{sb}\sigma^s\sigma^b sbV_{sb} = 0~~;~~t \in (t_{n-1}, t_n)
\label{expanded_7}
\end{align}
Define 
\begin{equation}
J(s,b,t_0^-) = \sup_{W^\prime} V(s,b,W^{\prime},t_0^-).
\label{expanded_equiv}
\end{equation}
It is then straightforward to see that formulation
(\ref{expanded_1}-\ref{expanded_7}) is equivalent to
problem~(\ref{PCEE_a}).\footnote{See \citet{forsyth_2019_c} for
discussion of a similar problem.}

We briefly describe our numerical solution approach. We refer the
reader to \citet{forsythlabahn2017} and \citet{Forsyth_2020_IME}
for further details. We start by solving the auxiliary problem
(\ref{expanded_1}-\ref{expanded_2}) with fixed values of $W^*$, $\kappa$
and $\alpha$. Since shorting of the stock index is not allowed, $S(t)
\geq 0$. We localize the domain $s>0$ on a finite localized domain
$ s \in [e^{\hat{x}_{\min}},e^{\hat{x}_{\max}}]$. The computational
domain for $s$ is discretized using $n_{\hat{x}}$ equally spaced
nodes in the $\hat{x} = \log s$ direction. Similarly, we define
the localized domain for $b>0$ to be $ b \in [b_{\min}, b_{\max}]
= [e^{y_{\min}},e^{y_{\max}}]$. The computational domain for $b>0$
is discretized using $n_y$ equally spaced nodes in the $y=\log b$
direction. Since the investor can become insolvent due to withdrawals,
we also define a mirror image grid for $b<0$ \citep{Forsyth_2020_IME}.

We use the Fourier methods described in \citet{forsythlabahn2017} to
solve PIDE (\ref{expanded_7}) between rebalancing times. Wrap-around
errors are minimized using the domain extension technique in \citet{forsythlabahn2017}.
The localized domain $[\hat{x}_{\min}, \hat{x}_{\max}]
= [\log(10^2) -8, \log(10^2)+8]$, with $[y_{\min}, y_{\max}] =
[\hat{x}_{\min}, \hat{x}_{\max}]$ (units for $e^{\hat{x}}$ are thousands of dollars). 
Numerical
tests showed that the errors involved in this domain localization
were at most in the fifth digit.

At rebalancing times, we discretize the equity fraction $p \in [0,1]$
using $n_y$ equally spaced nodes and evaluate the right hand side of
equation (\ref{expanded_3}) using linear interpolation. We then solve
the optimization problem (\ref{expanded_3}) using exhaustive search over
the discretized $p$ values.

Given an approximate solution of the auxiliary problem
(\ref{expanded_1}-\ref{expanded_2}) at $t=0$, which we denote by
$V(s, b, W^*, 0)$, we then compute the solution of problem
(\ref{PCEE_a}) using equation (\ref{expanded_equiv}). More specifically,
we solve
\begin{equation}
J(0,W_0,0^-) = \sup_{W^{\prime}} V(0, W_0, W^{\prime}, 0^-) 
\label{outer_opt}
\end{equation}
given initial wealth $W_0$.
We solve this outer optimization problem using
a one-dimensional optimization algorithm.\footnote{
Since the problem is not
guaranteed to be convex, we cannot be sure
that we converge to the  global maximum. Additional testing based on a
search over the finest grid suggests that we do indeed have the globally
optimal solution.}

If $W_t \gg W^*$ and $t \rightarrow T$, then \emph{Prob}$[ W_T < W^*]
\simeq 0$. In addition, for large values of $W_t$ the withdrawal is
capped at $q_{\max}$. As a result the objective function is almost
independent of the control, and thus determination of the control
becomes ill-posed. To avoid this, we change the objective function
(\ref{PCEE_a}) by adding a stabilizing term $\epsilon W_T$, giving
\begin{multline}
J(s,b,t_0^-) = \sup_{\mathcal{P}_0 \in \mathcal{A}}
               \sup_{W^*}
               \Biggl\{
               E_{\mathcal{P}_0}^{X_0^+,t_0^+}
               \Biggl[
               \sum_{i=0}^{i=M} q_i +
               \kappa\left( W^* + \frac{\min(W_T-W^*,0)}{\alpha}\right) +
               \epsilon W_T
               \Biggr.
               \Biggr. \\
               \Biggl.
               \Biggl.
               \bigg\vert X(t_0^-) = (s,b)
               \Biggr]
               \Biggr\}.
\label{stable_objective}
\end{multline}

A negative value for $\epsilon$ forces the strategy to invest in
the bond index when $W_t$ is very large and $t \rightarrow T$,
where the original control problem is ill-posed. This choice is
consistent with de-risking retirement assets as soon as possible
\citep{Merton_2014_HBR}. Setting $\epsilon = -10^{-4}$ gave the same
results as setting $\epsilon = 0$ to four digits for the summary
statistics of the problem solution. This is due to the fact that
outcomes with very large terminal wealth are highly unlikely.

\section{Data and Parameter Estimates}
\label{data_section}
As mentioned above, our model assumes that the retiree's portfolio is
allocated to either a stock index or a constant maturity bond index. In
order to have a long history encompassing expansions, recessions, stock
market booms and crashes, and different levels of interest rates, we
use US financial market data. In particular, the stock index is taken
to be the Center for Research in Security Prices (CRSP) Value-Weighted
Index\footnote{This is a total return index of the broad US stock
market, reflecting both distributions such as dividends and capital
gains/losses due to price changes.}, while the bond index is the
CRSP 30-Day Treasury bill (T-bill) Index. Both indexes are measured
on a monthly basis from January 1926 through December 2018, giving a
total of 1,116 observations. To work in real terms, we deflate both
indexes by the Consumer Price Index (CPI), which was also provided by
CRSP.\footnote{The CRSP data used in this study was obtained through
Wharton Research Data Services (WRDS). This service and the data
available thereon constitute valuable intellectual property and trade
secrets of WRDS and/or its third party suppliers.}

We use the threshold technique
\citep{mancini2009,contmancini2011,Dang2015a} to estimate
the parameters for the stochastic process models
(\ref{jump_process_stock}-\ref{jump_process_bond}) (see Appendix \ref{calibration_section}).
All estimated parameters reflect
real (inflation adjusted) returns. Table~\ref{fit_params} shows the annualized parameter
estimates. For reference, the table also gives the estimated
parameters for the two time series assuming geometric Brownian motion
(GBM).\footnote{The GBM parameter estimates are calculated using
maximum likelihood estimation.} For the threshold case, after removing any returns which occur at times
corresponding to jumps in either series, the correlation
$\rho_{sb}$ 
is then estimated
using the remaining sample
covariance.

The annualized real value-weighted stock index parameters in
Table~\ref{fit_params} for the double exponential jump diffusion model
correspond to an (uncompensated) drift rate of 8.6\% and a diffusive
volatility of 14.6\%. Jumps in the stock index are estimated to occur
about once every three years. Conditional on a jump occurring, a
downwards jump is about 3 times more likely than an upwards jump. The
mean jump size is about 23\% in the upward direction and 18\% in the
downward direction. Since the standard deviation is equal to the mean
for an exponentially distributed random variable, the magnitudes of
both upward and downward jumps can vary considerably. The corresponding
GBM parameter estimates imply a drift of about 8\% per annum, with a
volatility of 18.5\%. This volatility is higher than the diffusive
volatility for the jump model since in the GBM case this term
effectively combines the effects of volatility due to both diffusion and
jumps.

Turning to the T-bill index, the annualized jump model parameters
correspond to a real (uncompensated) drift of approximately 0.45\% and a
diffusive volatility of about 1.3\%. Jumps are estimated to occur about
every 2 years, slightly more often than for the stock index. Downward
jumps are again more likely than upward jumps, though somewhat less so
compared to the stock index. The mean jump size is around 1.5\% in the
upward direction, and about 1.7\% in the downward direction. The GBM
parameter estimates indicate a drift that is also about 0.45\%, and a
volatility of approximately 1.8\%. Finally, the correlation between the
diffusive terms for the two indexes is quite low, around .083 for the
jump model and .059 for the GBM case.

{\small
\begin{table}[tb]
\begin{center}
\begin{tabular}{cccccccc} \toprule
\multicolumn{8}{c}{Real CRSP Value-Weighted Stock Index} \\ \midrule
Method & $\mu^s$ & $\sigma^s$ & $\lambda^s$ & $p_{\text{\emph{up}}}^s$ &
  $\eta_1^s$ & $\eta_2^s$ & $\rho_{sb}$ \\ \midrule
Threshold ($\beta = 3$)
    & .08607 & .14600 & .32258 &  .23333 & 4.3578 & 5.5089 & .08311 \\
GBM & .08044 & .18460 &  N/A   &  N/A    &   N/A  &   N/A  & .05870 \\
  \midrule
\multicolumn{8}{c}{Real 30 Day T-bill Index} \\ \midrule
Method & $\mu^b$ & $\sigma^b$ & $\lambda^b$ & $p_{\text{\emph{up}}}^b$ &
  $\eta_1^b$ & $\eta_2^b$ & $\rho_{sb}$ \\ \midrule
Threshold ($\beta = 3$) 
    & .00454 & .01301 & .51610 & 0.39580 & 65.875 & 57.737 & .08311 \\
GBM & .00448 & .01814 & N/A    & N/A     & N/A    & N/A    & .05870 \\
  \bottomrule
\end{tabular}
\caption{Estimated annualized parameters for the double exponential jump
diffusion model~(\ref{jump_process_stock}-\ref{jump_process_bond}).
Sample period 1926:1 to 2018:12. GBM refers to a geometric Brownian
motion model (i.e.\ no jumps). The threshold method is described in
Appendix \ref{calibration_section}.
\label{fit_params}
}
\end{center}
\end{table}
}

\section{Investment Scenario}\label{investment_section}
In order to focus exclusively on decumulation, we consider an investor
just entering retirement at age 65 with savings of \$1 million. Our
investor is assumed to have the life expectancy characteristics of a
Canadian male. According to the CPM 2014 mortality table, this investor
has a 13\% probability of attaining the age of 95 and a 2\% probability
of reaching the century mark. We set the investment horizon $T$ to be 30
years.

We alter the standard ARVA spending rule so as to include an
annual floor of $q_{\min} = \$30,000$ and an annual cap of
$q_{\max}=\$80,000$. Recall that all quantities are expressed in real
(i.e.\ inflation-adjusted) terms. Our modified ARVA spending rule is
then
\begin{equation}
q_i = \max\left[q_{\min},\min\left(A(t_i)W_i^-, q_{\max}\right)\right]
\end{equation}
where $A(t_i)$ is given in equation (\ref{annuity_factor_1}). To
provide more context, a Canadian male who has worked for 40 years in a
high-earning occupation can expect to receive slightly over \$20,000 per
year in government benefits. Hence, we are assuming that the minimum
total amount needed per year is about $\$30,000+\$20,000=\$50,000$ per
year. Of course, the investor would like to withdraw more than the
minimum amount of \$30,000. However, as noted we also place a cap of
\$80,000 per year on withdrawals. The cap prevents the retiree from
reducing savings very quickly, establishing a buffer against potential
poor investment returns. We are thus effectively assuming that our
retiree has no need for income above $\$80,000+\$20,000=\$100,000$ per
year.\footnote{It is also worth noting that Canadian government benefits
are reduced when total income exceeds about \$80,000 per year, providing
further incentive to not withdraw more than the specified cap.}

Our retired investor withdraws cash and rebalances his portfolio at the
start of each year, beginning immediately. The interest rate used in the
ARVA calculation (\ref{annuity_factor_1}) is set equal to the estimated
value of $\mu_b$, which is given in Table~\ref{fit_params} as 0.454\%.
Table~\ref{base_case_1} summarizes the base case investment scenario.
Note that monetary units here and in the following tables and plots are
expressed in thousands of (real) dollars.

\begin{table}[tb]
\begin{center}
\begin{tabular}{lc} \toprule
Investment horizon $T$ (years) & 30  \\
Investor ($t=0$)   & 65-year old Canadian male \\
Mortality table    & CPM 2014 \\
Equity market index & CRSP value-weighted index (real) \\
Bond index & 30-day T-bill index (real) \\
Initial portfolio value $W_0$ & 1,000 \\
Cash withdrawal/portfolio rebalance times (years) & $t=0,1,\ldots, 30$ \\
$q_{\max}$           & 80 \\
$q_{\min}$           & 30 \\
Borrowing spread when $W_t < 0$ & $\mu_c^{b} = .02$ \\
Interest rate for ARVA computation (\ref{annuity_factor_1}) 
  & $\mu^b = 0.00454$ \\
Rebalancing interval (years) & 1  \\
Market parameters & See Table~\ref{fit_params} \\ \bottomrule
\end{tabular}
\caption{Base case input data. Monetary units: thousands of dollars.
The CPM 2014 mortality table is from the Canadian Institute of Actuaries.
\label{base_case_1}}
\end{center}
\end{table}

Since the investor uses a risky portfolio to fund minimum cash flows
annually, there is clearly no guarantee that he will not run out of
savings if he has survived to age 95. As outlined above, we seek
an investment strategy that minimizes risk as measured by expected
shortfall (ES), as defined by equation~(\ref{ES_def_1}). We use
$\alpha=5\%$, so we are trying to minimize the adverse consequences
measured by the average outcome in the worst 5\% of the distribution. As
indicated in Table~\ref{base_case_1}, when $W_t <0$ we assume that debt
accumulates at the rate given by the current return on 30-day T-bills
plus a spread of $\mu_c^b = 2\%$.

We focus solely on measured outcomes for the investment account, but
it is easy to imagine that our retiree also owns real estate such as
a home. In this case, the ES risk could be hedged using a reverse
mortgage with the home as collateral. However, we assume that the
investor wants to avoid using a reverse mortgage if at all possible, so
we seek an investment strategy that minimizes the magnitude of ES risk
on its own. Our scenario shares some features with the behavioural life
cycle approach originally described in \citet{Shefrin-Thaler:1988}. In
this framework, investors divide their wealth into separate ``mental
accounts'' containing funds intended for different purposes such as
current spending or future needs. The standard life cycle approach
assumes that wealth is completely fungible across any such accounts, so
that the same increase in wealth from any source (e.g.\ positive returns
for a financial market portfolio, an increase in the value of one's
house, lottery winnings, etc.) has the same effect on consumption. In
contrast, in the behavioral approach wealth is not completely fungible,
so the effects of increased wealth depend on the source of the increase.
In our case, even if the investor's wealth rises because the value of
his real estate has increased, there will be no impact on the amount
withdrawn from the retirement portfolio. The real estate account will
only be accessed as a last resort. It is assumed to be there in the
background if needed, but it is ignored in our analysis.

\section{Numerical Results: Synthetic Market}
\label{sec:synthetic_results}

We evaluate the performance of three alternative strategies based
on the scenario described by Table~\ref{base_case_1}: (i)~constant
withdrawals and investment portfolio rebalanced to maintain constant
asset allocation weights (in particular, we set $q_{\min}=q_{\max}=40$
instead of the values given in Table~\ref{base_case_1} so that
this strategy corresponds to the 4\% rule of \citet{Bengen1994});
(ii)~ARVA withdrawals as indicated in Table~\ref{base_case_1}
and investment portfolio rebalanced to maintain constant asset
allocation weights; and (iii)~ARVA withdrawals as indicated in
Table~\ref{base_case_1} and investment portfolio rebalanced to
optimal asset allocation weights, in accordance with solving the
pre-commitment EW-ES problem~(\ref{PCEE_a}) by the methods described in
Section~\ref{algo_section}. In each case, the performance evaluation
is based on Monte Carlo simulated paths of market returns based on the
parametric model (\ref{jump_process_stock}-\ref{jump_process_bond}),
with statistics of interest calculated across all paths. We refer to
this as a \emph{synthetic market}, since the data used is generated
by simulation of the parametric model rather than taken directly from
actual historical market returns.\footnote{We provide results based on
historical market returns below in Section~\ref{sec:historical_market}
and Appendix~\ref{detailed_historical}.}

We begin with the first strategy described above: constant withdrawals
based on the 4\% rule ($q_{\max}=q_{\min}=40$) and constant weights,
i.e.\ $p_{\ell} = \text{\emph{constant}}$ in equation~(\ref{PCEE_b}).
The results for the equity index weight $p_{\ell} = 0.0, 0.1, 0.2,
\dots, 1.0$ are shown in Table~\ref{const_wt_const_withdraw_30_day}.
This table also displays the results for $p_{\ell} = 0.15$, since this
is approximately the equity weight which results in the maximum ES. We
conjecture that this low equity weight is due to our use of ES to
measure risk, compared to the more typical standard deviation. As
$p_{\ell}$ increases past 0.15, the magnitude of ES increases strongly.
Taking on more equity market risk results obviously leads to higher
ES. Of course reward also rises, as shown by the median value of
terminal wealth $W_T$.\footnote{In general, our measure of reward is
total expected withdrawals. However, in this case the withdrawals
are fixed, so wealth is drawn down slowly given a sufficiently high
$p_{\ell}$ and decent equity market returns, resulting in relatively
high values for $W_T$.}

\begin{table}[tb]
\begin{center}
\begin{tabular}{cD{.}{.}{2}D{.}{.}{2}} \toprule
Equity Weight $p_{\ell}$ & \multicolumn{1}{c}{ES ($\alpha=5\%$)} & 
  \multicolumn{1}{c}{Median[$W_T$]} \\ \midrule
0.00 & -344.95 & -192.14 \\
0.10 & -284.46 & -55.17 \\
0.15 & -284.28 & 22.29  \\
0.20 & -294.32 & 108.70 \\
0.30 & -332.05 & 310.12 \\
0.40 & -384.62 & 550.25 \\
0.50 & -447.55 & 828.81 \\
0.60 & -518.24 & 1143.18 \\
0.70 & -594.67 & 1490.44 \\
0.80 & -675.08 & 1862.64 \\
0.90 & -758.57 & 2249.94 \\
1.00 & -844.37 & 2637.77 \\ \bottomrule
\end{tabular}
\caption{Synthetic market results for constant withdrawals with constant
weights, i.e.\ assuming the scenario from Table~\ref{base_case_1} except
that $q_{\max} = q_{\min} = 40$ and $p_{\ell} = \text{\emph{constant}}$
in equation~(\ref{PCEE_b}). Units: thousands of dollars. Statistics are
based on $2.56 \times 10^6$ Monte Carlo simulated paths.
\label{const_wt_const_withdraw_30_day}}
\end{center}
\end{table}

To see the benefit of the ARVA withdrawal strategy, we repeat
the Monte Carlo simulations from above, except that here the
ARVA spending strategy (\ref{annuity_factor_1}) is used with the
constraints $q_{\min} = 30$ and $q_{\max}=80$. The results are
shown in Table~\ref{const_wt_ARVA_30_day}, which has an additional
column compared to Table~\ref{const_wt_const_withdraw_30_day}.
This extra column shows the expected average withdrawals over the
decumulation period, $\text{EW}/(M+1) = \sum_i q_i/M$.\footnote{This
column was excluded from Table~\ref{const_wt_const_withdraw_30_day}
because in that case the annual withdrawals were constant
at 40.} In Table~\ref{const_wt_ARVA_30_day} the largest ES
is $-38.43$ for $p_{\ell}=0.2$. This equity weight gives an
expected annual withdrawal of $42.07$. Recall that the largest
ES from Table~\ref{const_wt_const_withdraw_30_day} was $-284$,
with constant annual withdrawals of 40. There is a dramatic
improvement in ES, despite higher average withdrawals. As another
observation, in Table~\ref{const_wt_ARVA_30_day} the strategy
with $p_{\ell}=0.7$ has better ES than the best result in
Table~\ref{const_wt_const_withdraw_30_day}, while the average expected
withdrawal is $59.13$, again compared to the constant withdrawal of
$q=40$. Overall, our comparison between strategies with constant asset
weights and constant vs.\ variable spending (the ARVA rule augmented
with a floor and a cap) is consistent with the results in studies such
as \citet{Pfau_2015}, albeit with different measures of risk and reward:
a variable spending rule allows for both higher average withdrawals and
lower risk as measured by ES.

\begin{table}[tb]
\begin{center}
\begin{tabular}{cD{.}{.}{2}D{.}{.}{2}D{.}{.}{2}} \toprule
Equity Weight $p_{\ell}$ & \multicolumn{1}{c}{ES ($\alpha=5\%$)} & 
  \multicolumn{1}{c}{$\text{EW}/(M+1)$} &
  \multicolumn{1}{c}{Median[$W_T$]} \\ \midrule
0.0 & -78.89  & 34.80 & -12.36 \\
0.1 & -39.60  & 37.85 &  31.48 \\
0.2 & -38.43  & 42.07 &  64.31 \\
0.3 & -54.01  & 46.95 &  90.01 \\
0.4 & -82.92  & 51.46 &  111.32 \\
0.5 & -124.19 & 54.95 &  138.11 \\
0.6 & -176.92 & 57.42 &  179.68 \\
0.7 & -239.69 & 59.13 &  275.02 \\
0.8 & -310.78 & 60.30 &  486.56 \\
0.9 & -387.96 & 61.07 &  739.74 \\
1.0 & -469.67 & 61.56 &  1013.85 \\ \bottomrule
\end{tabular}
\caption{Synthetic market results for ARVA withdrawals with constant
weights, i.e.\ assuming the scenario from Table~\ref{base_case_1} except
that $p_{\ell}=\text{\emph{constant}}$ in equation~(\ref{PCEE_b}). There
are $M=30$ rebalancing dates and $M+1$ withdrawals. Units: thousands
of dollars. Statistics are based on $2.56 \times 10^6$ Monte Carlo
simulated paths.
\label{const_wt_ARVA_30_day}}
\end{center}
\end{table}

We next consider our third strategy of ARVA withdrawals with optimal
asset allocation. In particular, we consider the scenario described
in Table~\ref{base_case_1} and solve for the optimal control $p(W,t)$
for the pre-commitment EW-ES problem given by equation~(\ref{PCEE_a})
using the methods discussed in Section~\ref{algo_section}. We store the
optimal control and then carry out Monte Carlo simulations to calculate
statistical properties as above but with applying $p(W,t)$ along each
path rather than rebalancing to constant weights. We reiterate that
for all times $t>0$, this corresponds to the induced time consistent
strategy that solves equation~(\ref{timec_equiv}).

Before presenting the main results, we first verify the convergence
of the algorithm given in Section \ref{algo_section} that is used to
solve the optimal control problem given by equation~(\ref{PCEE_a}).
Table~\ref{conservative_accuracy} shows a test with various levels
of grid refinement for a fixed value of $\kappa = 2.5$ in equation
(\ref{PCEE_a}). At each grid refinement, we compute and store the
optimal controls which are then used in Monte Carlo simulations. The
algorithm in Section \ref{algo_section} and the Monte Carlo simulations
are in good agreement. As expected, the value function appears to be
converging at almost a quadratic rate. The other quantities ES and
expected average withdrawals which are derived from the algorithm in
Section~\ref{algo_section} converge a bit more erratically. Results
reported below for all cases with optimal asset allocation are
calculated using the finest grid from Table~\ref{conservative_accuracy}.

{\small
\begin{table}[tb]
\begin{center}
\begin{tabular}{cD{.}{.}{3}D{.}{.}{4}D{.}{.}{4}D{.}{.}{3}D{.}{.}{4}} \toprule
\multicolumn{4}{c}{Algorithm in Section~\ref{algo_section}} &
\multicolumn{2}{c}{Monte Carlo} \\ \cmidrule(lr{1ex}){1-4} 
  \cmidrule(lr{1ex}){5-6}
 & \multicolumn{2}{c}{\mbox{}} & \multicolumn{1}{c}{Value} &
   \multicolumn{2}{c}{\mbox{}} \\
Grid & \multicolumn{1}{c}{ES ($\alpha=5\%$)} &
  \multicolumn{1}{c}{$\text{EW}/(M+1)$} &
  \multicolumn{1}{c}{Function} &
  \multicolumn{1}{c}{ES ($\alpha=5\%$)} &
  \multicolumn{1}{c}{$\text{EW}/(M+1)$} \\ \cmidrule(lr{1ex}){1-4}
  \cmidrule(lr{1ex}){5-6}
$512 \times 512$   & -64.633 & 54.8128 & 1537.6144 & -59.326 & 54.779 \\ 
$1024 \times 1024$ & -61.305 & 54.8377 & 1546.5833 & -59.381 & 54.802 \\
$2048 \times 2048$ & -60.196 & 54.8230 & 1549.0359 & -59.469 & 54.812 \\
 \bottomrule
\end{tabular}
\caption{Convergence test for the algorithm from
Section~\ref{algo_section} used to determine the optimal
asset allocation strategy to solve the pre-commitment EW-ES
problem~(\ref{PCEE_a}) with $\kappa=2.5$ for the scenario from
Table~\ref{base_case_1}. The Monte Carlo method used $2.56 \times 10^6$
simulated paths. The grid is reported as $n_x \times n_b$, where $n_x$
is the number of nodes in the $\log s$ direction and $n_b$ is the number
of nodes in the $\log b$ direction. There are $M=30$ rebalancing dates
and $M+1$ withdrawals. Units: thousands of dollars. The value of $W^*$
in equation~(\ref{PCEE_a}) is 4.13 on the finest grid.
\label{conservative_accuracy}}
\end{center}
\end{table}
}

Table \ref{optimal_wt_ARVA_30_day} shows the results for
the ARVA spending rule with optimal asset allocation from
solving the pre-commitment EW-ES problem~(\ref{PCEE_a}) for
various values of $\kappa$. In addition to ES, expected
average withdrawals $\text{EW}/{(M+1)}$, and median $W_T$,
Table~\ref{optimal_wt_ARVA_30_day} shows the average throughout the
investment horizon of the median value of the fraction of the portfolio
invested in equities in the furthest right column. This gives a rough
indication of the equity market risk taken on over the period. As
indicated by equation~(\ref{eq:informal}), increasing $\kappa$ places
more emphasis on risk relative to reward. As a result, the optimal
equity allocation tends to decrease with $\kappa$. This is also
reflected in reduced median $W_T$ and expected average withdrawals. The
benefit from higher $\kappa$ is a lower magnitude of ES. Consider the
case here with $\kappa = 5$ which results in ES of $-37.91$, expected
average withdrawals of 52.35, and median $W_T$ of 129.97. This strategy
has an average median equity allocation of 0.34. Contrast this with the
result reported in Table~\ref{const_wt_ARVA_30_day} for $p_{\ell}=0.2$,
which had about the same ES ($-38.43$), but expected average withdrawals
of just 42.07 and median terminal wealth of 64.31. In this case, using
an optimal asset allocation strategy compared to a constant weight
strategy results in about the same ES but significantly higher average
withdrawals and about twice as much median $W_T$. This attests to the
benefits of optimizing the asset allocation strategy, in addition to
allowing for variable withdrawals.

\begin{table}[tb]
\begin{center}
\begin{tabular}{rD{.}{.}{2}D{.}{.}{2}D{.}{.}{2}D{.}{.}{3}} \toprule
\multicolumn{1}{c}{$\kappa$} & \multicolumn{1}{c}{ES ($\alpha=5\%$)} &
  \multicolumn{1}{c}{$\text{EW}/(M+1)$} &
  \multicolumn{1}{c}{Median[$W_T$]} &
  \multicolumn{1}{c}{$\sum_i \text{Median}(p_i)/M$} \\ \midrule
0.1    & -459.93 & 63.01 & 266.43 & .455 \\
0.3    & -308.26 & 61.67 & 258.64 & .458 \\
0.5    & -209.63 & 60.15 & 250.59 & .451 \\
1.0    & -119.10 & 57.91 & 237.06 & .416 \\
1.75   & -77.02  & 56.04 & 208.67 & .390 \\
2.5    & -59.47  & 54.81 & 180.36 & .375 \\
5.0    & -37.91  & 52.35 & 129.97 & .340 \\
10.0   & -25.90  & 49.59 & 93.19  & .291 \\
20.0   & -19.78  & 46.82 & 66.53  & .243 \\ 
100.0  & -15.98  & 42.35 & 44.77  & .173 \\
1000.0 & -15.74  & 40.30 & 39.52  & .139 \\ \bottomrule
\end{tabular}
\caption{Synthetic market results for ARVA withdrawals with optimal
asset allocation based on the scenario from Table~\ref{base_case_1}
for various values of $\kappa$. The optimal control that solves the
pre-commitment EW-ES problem~(\ref{PCEE_a}) is computed using the
algorithm given in Section~\ref{algo_section}, stored, and then applied
in the Monte Carlo simulations. There are $M=30$ rebalancing dates and
$M+1$ withdrawals. Units: thousands of dollars. Statistics are based
on $2.56 \times 10^6$ Monte Carlo simulated paths. The stabilization
parameter in equation~(\ref{stable_objective}) is $\epsilon = -10^{-4}$.
\label{optimal_wt_ARVA_30_day}}
\end{center}
\end{table}

\begin{figure}[tb]
\centerline{%
\begin{subfigure}[t]{.50\linewidth}
\centering
\includegraphics[width=\linewidth]{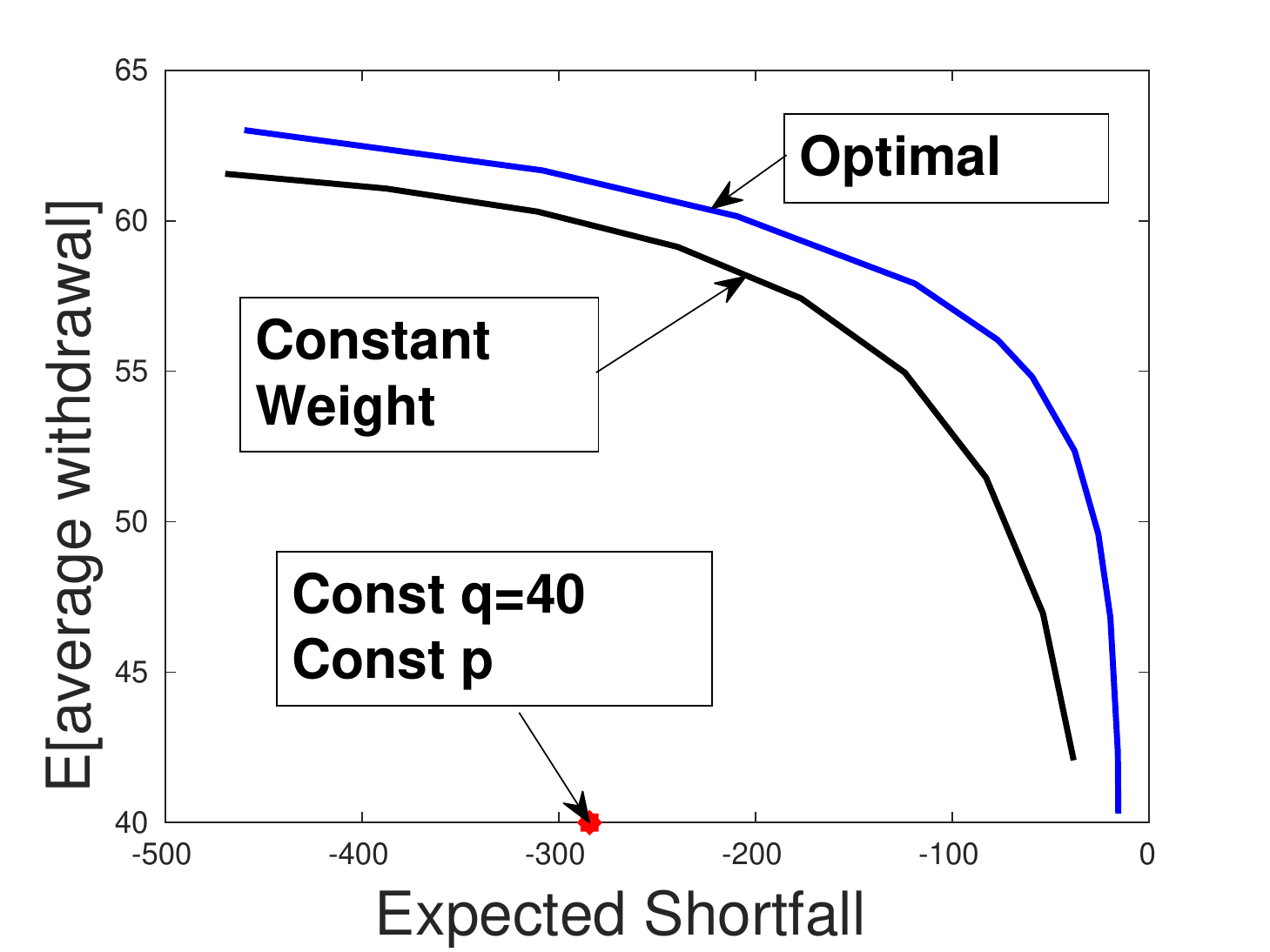}
\caption{ARVA withdrawals with optimal and constant weight
asset allocation, and the single best point for a
constant withdrawal strategy with $q=40$ and constant
weight asset allocation. For this point, $p_{\ell}=0.15$.}
\label{compare_frontier_fig}
\end{subfigure}
\begin{subfigure}[t]{.50\linewidth}
\centering
\includegraphics[width=\linewidth]{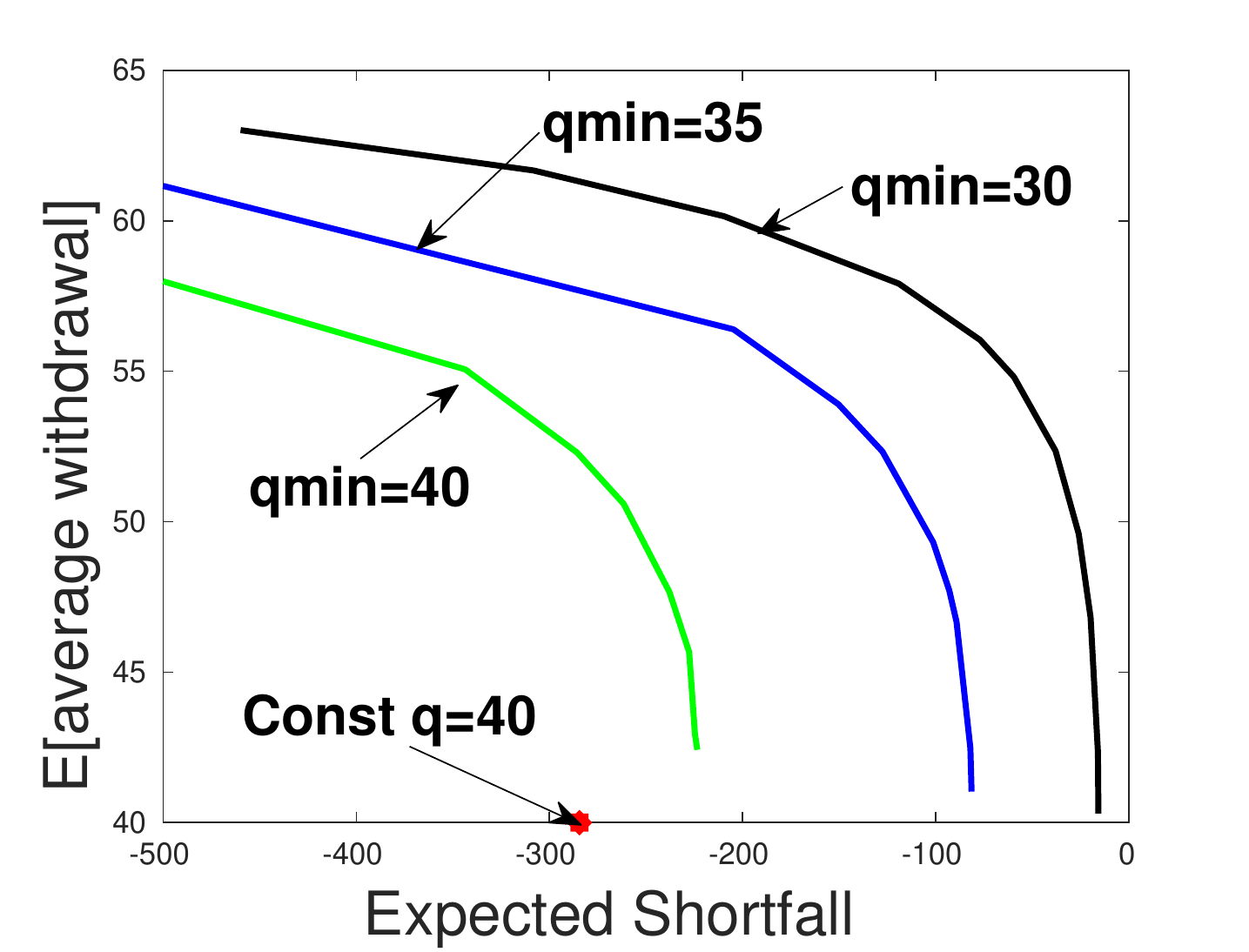}
\caption{ARVA withdrawals with optimal asset allocation with
$q_{\max}=80$ and various values for $q_{\min}$, and the single
best point for a constant withdrawal strategy with $q=40$ and
constant weight asset allocation. For this point, $p_{\ell}=0.15$.}
\label{compare_frontier_qmin_30_fig}
\end{subfigure}
}
\caption{Efficient frontiers in the synthetic market for the scenario
from Table~\ref{base_case_1}. All non-Pareto points have been 
removed. Units: thousands of dollars.}
\label{frontier_synthetic_fig}
\end{figure}

To further investigate the benefits of using an optimal
asset allocation strategy, we plot the efficient frontiers
of expected average withdrawals $\text{EW}/(M+1)$ vs.\ ES in
Figure~\ref{compare_frontier_fig}. We show these frontiers for (i)
the ARVA spending rule with optimal asset allocation as computed
by solving the pre-commitment EW-ES problem (\ref{PCEE_a}), with
results provided in Table~\ref{optimal_wt_ARVA_30_day}; (ii)
the ARVA spending rule with a constant weight asset allocation
strategy, with results shown in Table~\ref{const_wt_ARVA_30_day};
and (iii) a constant withdrawal of $q=40$ with a constant weight
strategy, with just the best result (i.e.\ highest ES) from
Table~\ref{const_wt_const_withdraw_30_day}.\footnote{This last
case leads to just a single point in our plot since withdrawals
are fixed at 40 regardless of the asset allocation and all other
constant equity weights lead to lower ES.} Note that we have removed
all non-Pareto points from these frontiers for plotting purposes.
Figure~\ref{compare_frontier_fig} shows that even with constant asset
allocation weights the ARVA spending rule is much more efficient than a
constant withdrawal strategy which also has constant asset allocation
weights. In fact, ARVA alone provides about 50\% higher expected
average withdrawals for the same ES achieved by a constant withdrawal
strategy by allowing for a higher stock allocation and limited income
variability. The case with optimal asset allocation with the ARVA
spending rule plots above the corresponding case with constant asset
allocation, with a larger gap between them for higher values of
ES.

%\begin{figure}[tb]
%\begin{center}
%\includegraphics[width=3in]{compare_frontier}
%\caption{Efficient frontiers in the synthetic market for ARVA
%withdrawals with optimal asset allocation and ARVA withdrawals
%with constant weight asset allocation for the scenario given in
%Table~\ref{base_case_1}. Also shown is the single best point for a
%constant withdrawal strategy having $q=40$ with constant weight asset
%allocation (for this point, $p_{\ell}=0.15$). All non-Pareto points
%have been removed. Units: thousands of dollars.
%\label{compare_frontier_fig}}
%\end{center}
%\end{figure}

To see the impact of the minimum required withdrawals,
Figure~\ref{compare_frontier_qmin_30_fig} displays efficient
frontiers for the ARVA spending rule with optimal asset allocation
for various values of $q_{\min}$, keeping $q_{\max}=80$. As a point
of comparison, we also show the point corresponding to the constant
weight strategy with $p_{\ell}=0.15$, which gives the highest ES for
constant withdrawals of $q=40$. As $q_{\min}$ rises  the efficient
frontiers move down and to the left, as expected. However, even for
$q_{\min} = 40$, the efficient frontier plots well above the best
point for constant withdrawals of $q=40$ with constant asset weights.
This indicates that much larger expected average withdrawals can
be attained at no cost in terms of higher ES through the use of
the ARVA spending rule and optimal asset allocation. Surprisingly,
Figure~\ref{compare_frontier_qmin_30_fig} shows that the combination
of ARVA and optimal control increases EW by 25\%, even when income is
constrained to be no less than for the constant withdrawal case.

%\begin{figure}[tb]
%\begin{center}
%\includegraphics[width=3.0in]{frontier_qmin_35}
%\caption{Efficient frontiers in the synthetic market for ARVA
%withdrawals with optimal asset allocation for the scenario given
%in Table~\ref{base_case_1} with $q_{\max}=80$ and different values
%of $q_{\min}$. Also shown is the single best point for a constant
%withdrawal strategy having $q=40$ with constant weight asset allocation
%(for this point, $p_{\ell}=0.15$). All non-Pareto points have been
%removed. Units: thousands of dollars.
%\label{compare_frontier_qmin_30_fig}}
%\end{center}
%\end{figure}

\begin{figure}[htb]
\centerline{%
\begin{subfigure}[t]{.33\linewidth}
\centering
\includegraphics[width=\linewidth]{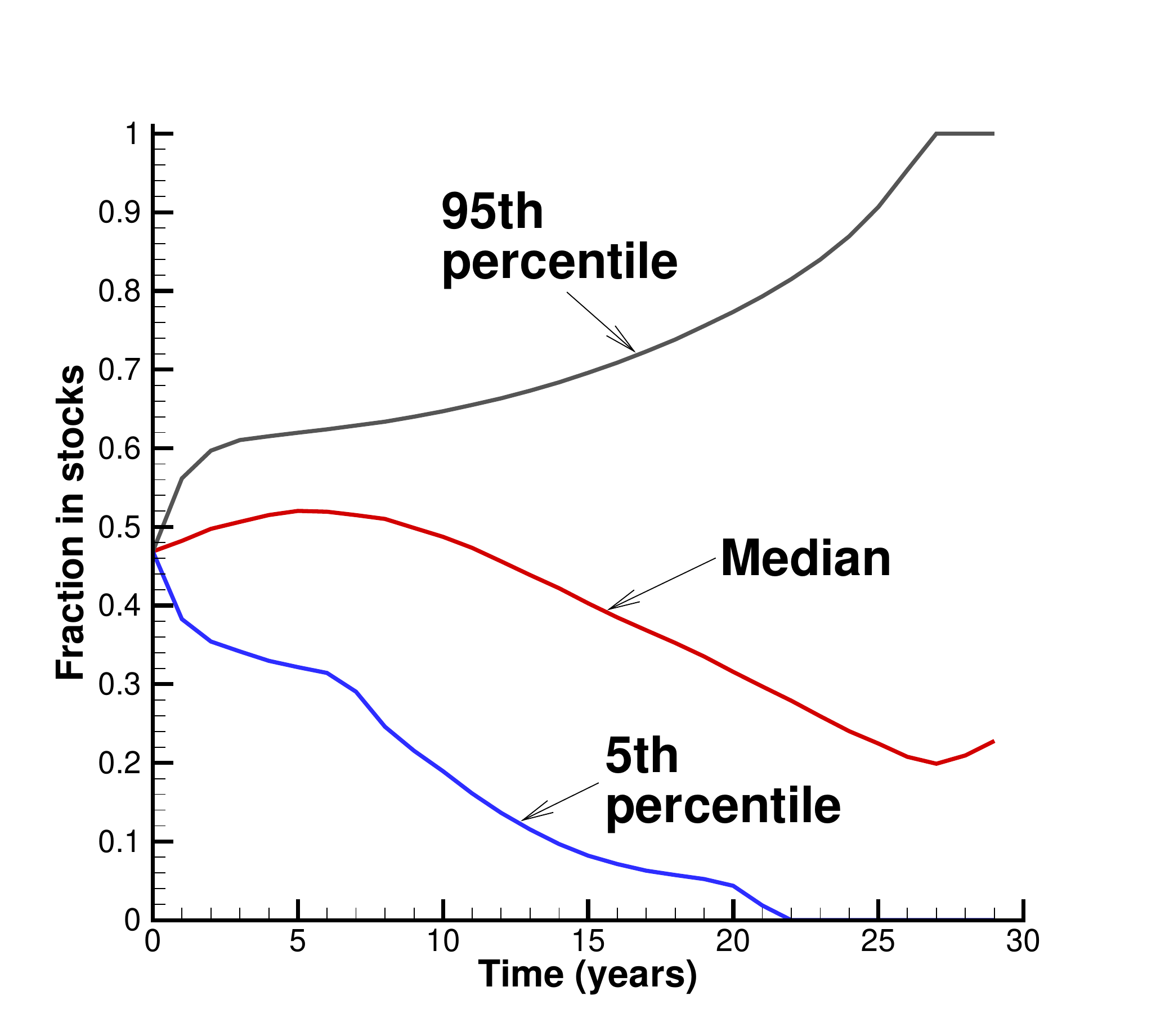}
\caption{Percentiles of the fraction invested in the stock index.}
\label{percentile_stocks_fig}
\end{subfigure}
\begin{subfigure}[t]{.33\linewidth}
\centering
\includegraphics[width=\linewidth]{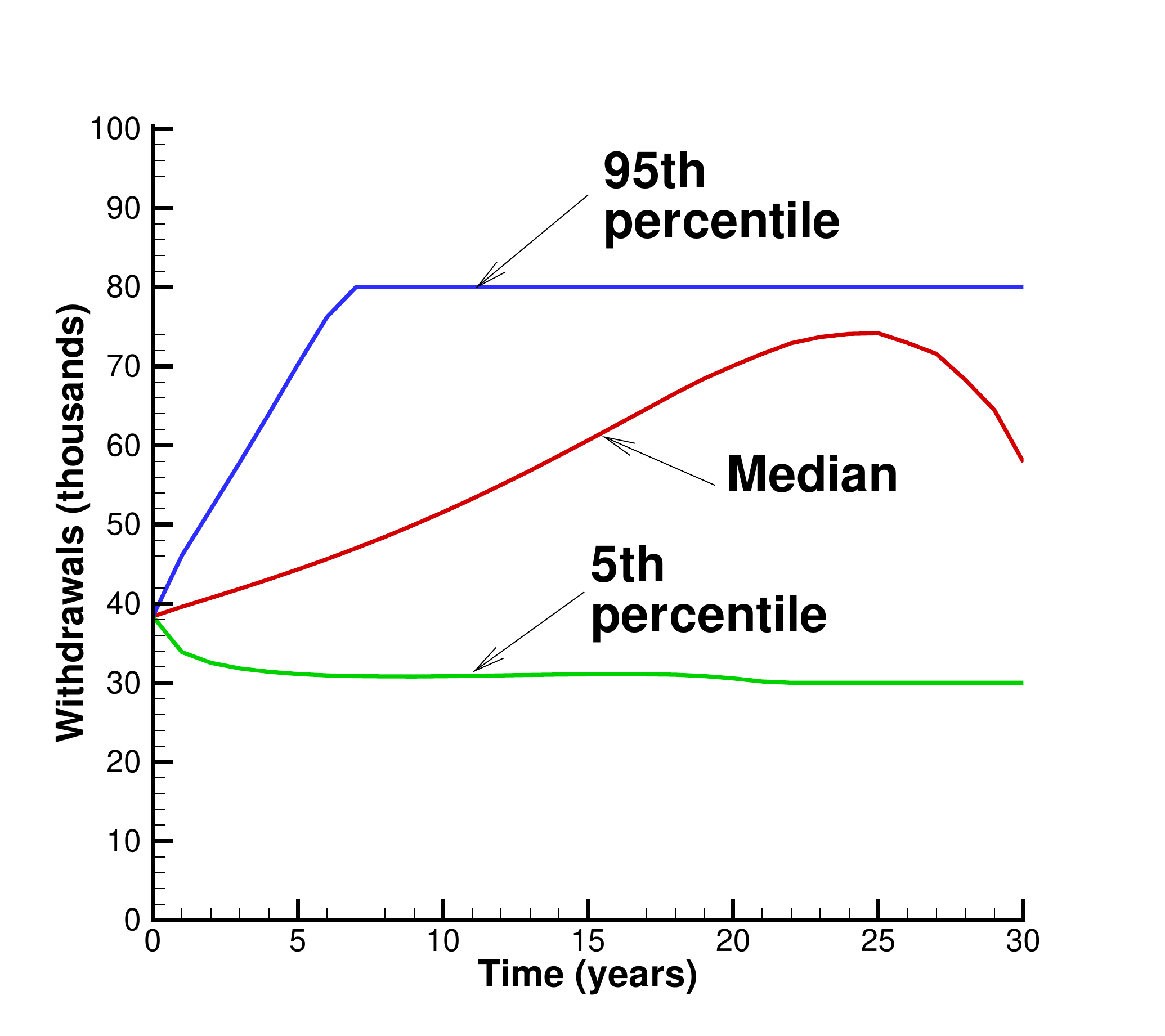}
\caption{Percentiles of withdrawals.}
\label{percentile_q_fig}
\end{subfigure}
\begin{subfigure}[t]{.33\linewidth}
\centering
\includegraphics[width=\linewidth]{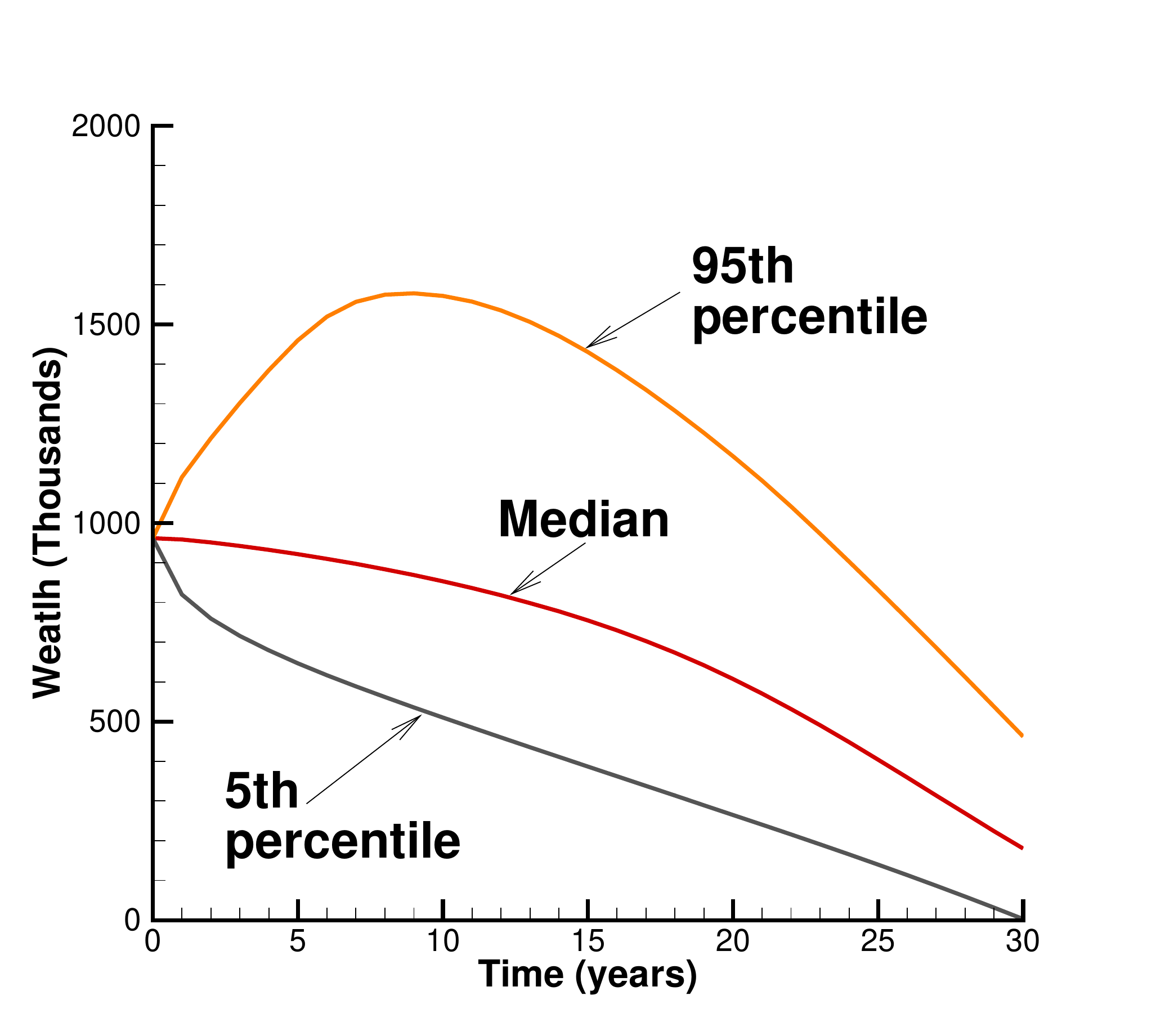}
\caption{Percentiles of wealth.}
\label{percentile_wealth_fig}
\end{subfigure}
}
\caption{Percentiles in the synthetic market of the fraction
invested in the stock index, withdrawals, and wealth for the scenario
from Table~\ref{base_case_1} with ARVA withdrawals and optimal asset
allocation. Based on $2.56 \times 10^6$ Monte Carlo simulated paths.
Units: thousands of dollars.
\label{percentiles_synthetic}}
\end{figure}

Additional insight into the properties of the ARVA spending rule in
conjunction with an optimal asset allocation strategy can be gleaned
from Figure \ref{percentiles_synthetic} showing the 5th, 50th, and 95th
percentiles of the fraction of the retiree's portfolio invested in the
stock index, withdrawals, and wealth throughout the 30-year decumulation
period. The optimal controls are computed by solving the pre-commitment
EW-ES problem~(\ref{PCEE_a}) with $\kappa=2.5$ and then used in Monte
Carlo simulations to generate these plots. The general trend is for the
equity index weight to decline over time, but there are cases where it
rises significantly instead. Median withdrawals increase for the first
25 years, before falling off a bit. The 5th percentile of withdrawals
quickly drops to $q_{\min}=30$ and remains there. On the other hand,
the 95th percentile of withdrawals rises sharply for about the first 5
years, and then stays at $q_{\max}=80$. Median wealth trends downward
consistently over time, as does the 5th percentile of wealth. The 95th
percentile of wealth rises over the first several years, before also
falling off fairly sharply.

\begin{figure}[tb]
\begin{center}
\includegraphics[width=3.5in]{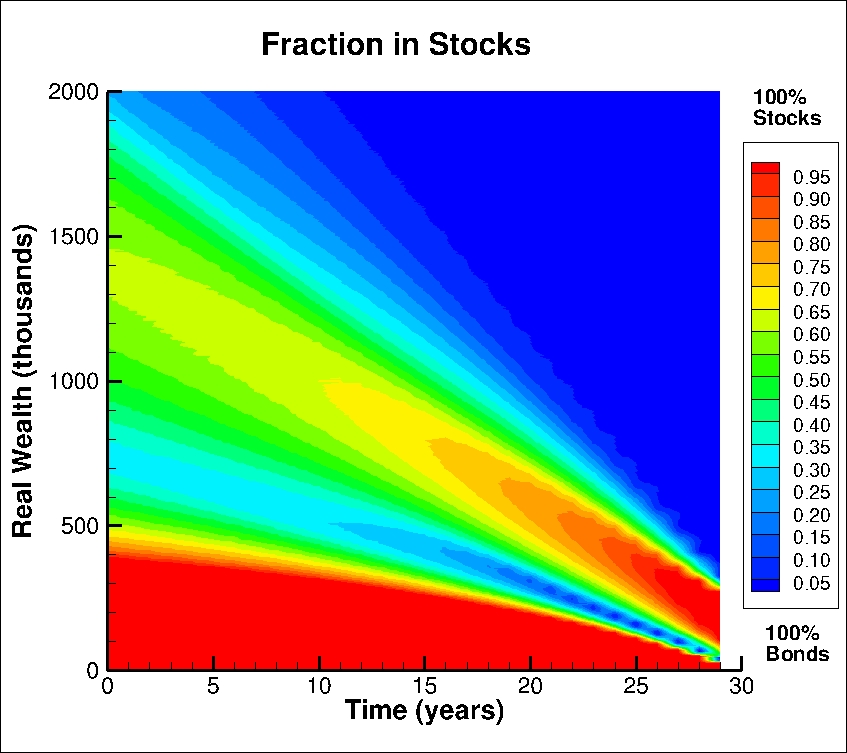}
\caption{Heat map of controls computed from solving the pre-commitment
EW-ES problem~(\ref{PCEE_a}) for $\kappa=2.5$ with ARVA withdrawals
based on the scenario from Table~\ref{base_case_1}. The
stabilization parameter in equation~(\ref{stable_objective}) is
$\epsilon=-10^{-4}$.
\label{heat_map_fig}}
\end{center}
\end{figure}

Recall that Proposition \ref{equiv_thm} states that the solution of
the pre-commitment EW-ES problem~(\ref{PCEE_a}) has the same controls at
time zero as the induced time consistent problem~(\ref{timec_equiv}).
Given any point in $(W_{t_n}, t)$ space ($t_n$ are the rebalancing
times), maximizing
\begin{equation}
\tilde{J}(s,b,t_n^-) = \sup_{\mathcal{P}_n \in \mathcal{A}}
  \Biggl\{E_{\mathcal{P}_n}^{X_n^+,t_n^+}
  \Biggl[\sum_{i=1}^{i=M}q_i + \frac{\kappa\min(W_T-\mathcal{W}^*,0)}{\alpha}
  + \epsilon W_T \bigg\vert X(t_n^-)=(s,b)
  \Biggr]
  \Biggr\}
\label{timec_equiv_2}
\end{equation}
leads to the optimal strategy depicted in the heat map contained in
Figure~\ref{heat_map_fig}. For this example, if we set $\kappa = 2.5$
in problem~(\ref{PCEE_a}), then $W^* = 4.13$. Recall that $W^*$ is set
to be the value such that \emph{Prob}$[ W_T < W^*] = \alpha$
\emph{as determined at time zero}.\footnote{In all of our examples, we 
maximize ES at the $\alpha=.05$ level.}

The structure of the heat map can be understood as follows. As
$t\rightarrow T$, there are multiply-connected regions of all bond
and all stock portfolios. For small values of wealth, the optimal strategy
is to be fully invested in stocks, thus attempting to
maximize ES. As wealth increases, \emph{Prob}$[W_T < W^*]$ is small, and
the investor switches to a portfolio that is heavily weighted towards
the bond index to protect against the ES risk. If wealth
increases further, the investor moves to investing more in stocks,
in order to maximize withdrawals. Finally, for large values of wealth,
there is little chance that $W_T < W^*$. Since the withdrawals
are capped at 80 per year, there is no incentive to take on
any more risk. In this case, the stabilization term $\epsilon W_T$ in
equation (\ref{timec_equiv_2}) comes into effect. Since $\epsilon =
-10^{-4} < 0$, this forces the strategy back into bonds.

It is useful to examine Figure \ref{heat_map_fig} with reference to the
median wealth shown in Figure \ref{percentile_wealth_fig}. The initial
wealth of $1000$ is in the green region, with equity weight $\simeq
0.50$. As $t \rightarrow T$, the optimal control attempts to guide real
wealth into the sweet spot between the lower blue zone and the upper red
zone. The lower blue zone then acts as a barrier to lower wealth (i.e.\
running out of cash), since the portfolio becomes very stable with a
large fraction of bonds. Above the lower blue zone, the allocation can
vary considerably in an effort to maximize the total withdrawals,
especially with a short time remaining.

Figure \ref{heat_map_fig} also shows the effect of different starting
values of wealth $W_0$, keeping a minimum withdrawal of $q_{\min} = 30$.
For example, with $W_0=400$ the investor has no choice but to start
with an investment of 100\% in stocks and hope for the best. This is
essentially a ``Hail Mary'' strategy, with little chance of success.
On the other hand, if $W_0=2000$, the investor will start off being
completely invested in bonds with very high probability of success.

\section{Numerical Results: Historical Market}
\label{sec:historical_market}

We continue to compute and store the optimal controls based on the
parametric model (\ref{jump_process_stock}-\ref{jump_process_bond})
as in the synthetic market case.  As a robustness test, we now calculate statistics
using these stored controls, but with bootstrapped historical real
return data rather than Monte Carlo simulations following the
parametric model. We employ  the stationary block bootstrap method
\citep{politis1994,politis2004} to generate many bootstrap simulated
paths. A single path entails sampling randomly sized blocks from the
historical data with replacement and pasting them together to cover
the entire decumulation period of $T=30$ years.\footnote{Sampling in
blocks helps to incorporate any serial correlation that is present in
the data.} The blocksize is generated randomly according to a geometric
distribution with expected blocksize $\hat{b}$, which helps to mitigate
the effects of a fixed block size.

We implement an algorithm from \citet{politis2009} to determine the
optimal expected blocksize $\hat{b}$ for the bond and stock indexes
separately. This indicates that the optimal expected blocksizes are
0.25 and 4.2 years for the stock and bond indexes respectively.
However, to allow for possible contemporaneous dependence between the
two indexes we use paired sampling to simultaneously draw returns
from both series. Given the large difference in optimal expected
blocksize for the two indexes, it is not obvious what should be done
for paired sampling. One possibility is to use an average of the
two estimates, suggesting about 2 years. We do this, but we also
give results for a range of expected blocksizes as a robustness
check.\footnote{Detailed pseudo-code for block bootstrap resampling can
be found in \citet{Forsyth_Vetzal_2019a}.}

In these bootstrap simulations, we continue to use the average
historical real (uncompensated) drift for the T-bill index $\mu^b$ as
the interest rate in the ARVA computation (\ref{annuity_factor_1}). This
avoids the problem of fluctuating withdrawal amounts which are driven
just by the bootstrap resampling methods. It is also a conservative
approach since $\mu^b \simeq 0$.

We first explore the effect of the expected blocksize $\hat{b}$.
Table~\ref{optimal_wt_ARVA_30_day_boot_kappa} shows the results
computed by solving the pre-commitment EW-EW problem~(\ref{PCEE_a})
in the synthetic market with $\kappa=2.5$ and then using this control
with block bootstrap resampling having various expected blocksizes
$\hat{b}$. For ease of comparison, the table also provides the results
for $\kappa=2.5$ in the synthetic market that were previously shown in
Table~\ref{optimal_wt_ARVA_30_day}. The historical market results in
Table~\ref{optimal_wt_ARVA_30_day_boot_kappa} are generally similar
to the corresponding synthetic market result, at least for values of
$\hat{b}$ between 0.5 and 2 years. The reported ES values for the
historical market are consistently a bit better than in the synthetic
market, while expected average withdrawals and median terminal wealth
are quite comparable. However, the average of the median value of the
equity weight is a bit higher, clustering at or above 0.4 for the
historical market compared to 0.375 for the synthetic market. Results
reported below use $\hat{b}=2$ years, as this is (approximately) the
average of the optimal expected blocksizes for the two indexes.

\begin{table}[tb]
\begin{center}
\begin{tabular}{rD{.}{.}{2}D{.}{.}{2}D{.}{.}{2}D{.}{.}{2}} \toprule
\multicolumn{1}{c}{$\hat{b}$} &
\multicolumn{1}{c}{ES ($\alpha=5\%$)} &
\multicolumn{1}{c}{$\text{EW}/(M+1)$} &
\multicolumn{1}{c}{$\text{Median}[W_T]$} &
\multicolumn{1}{c}{$\sum_i \text{Median}(p_i)/M$} \\ \midrule
\multicolumn{5}{c}{Synthetic Market (from Table~\ref{optimal_wt_ARVA_30_day})}
 \\ \midrule
N/A & $-59.47$ & 54.81 & 180.36 & .375 \\ \midrule
\multicolumn{5}{c}{Historical Market} \\ \midrule
0.25 years & $-43.93$ & 54.66 & 169.98 & .398 \\
0.5 years  & $-53.47$ & 54.88 & 174.49 & .400 \\
1 year     & $-50.83$ & 55.07 & 178.59 & .407 \\
2 years    & $-40.80$ & 55.15 & 180.32 & .416 \\
5 years    & $-26.53$ & 55.14 & 182.19 & .420 \\ \bottomrule
\end{tabular}
\caption{Historical market results for ARVA withdrawals with optimal
asset allocation based on the scenario from Table~\ref{base_case_1}
for various expected blocksizes $\hat{b}$. The optimal control that
solves the pre-commitment EW-ES problem~(\ref{PCEE_a}) is computed
using the algorithm given in Section~\ref{algo_section}, stored, and
then applied to bootstrap resamples of the monthly data from 1926:1
to 2018:12. Statistics are based on $10^5$ bootstrapped paths. There
are $M=30$ rebalancing dates and $M+1$ withdrawals. The scalarization
parameter in equation~(\ref{PCEE_a}) is $\kappa=2.5$ and the stabilization
parameter in equation~(\ref{stable_objective}) is $\epsilon = -10^{-4}$.
Units: thousands of dollars.
\label{optimal_wt_ARVA_30_day_boot_kappa}}
\end{center}
\end{table}

\begin{figure}[tb]
\centerline{%
\begin{subfigure}[t]{.33\linewidth}
\centering
\includegraphics[width=\linewidth]{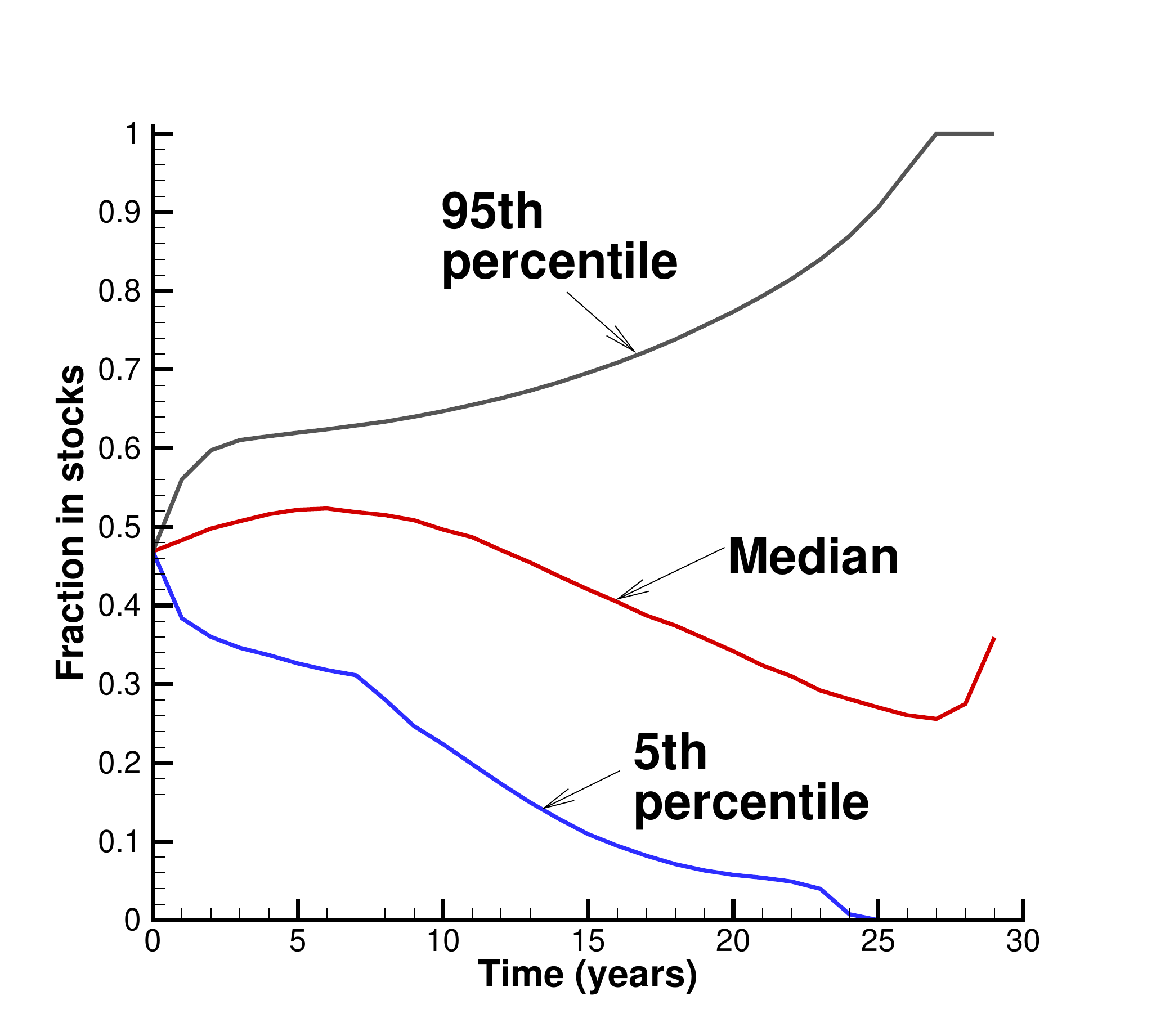}
\caption{Percentiles of the fraction invested in the stock index.}
\label{percentile_stocks_boot_fig}
\end{subfigure}
\begin{subfigure}[t]{.33\linewidth}
\centering
\includegraphics[width=\linewidth]{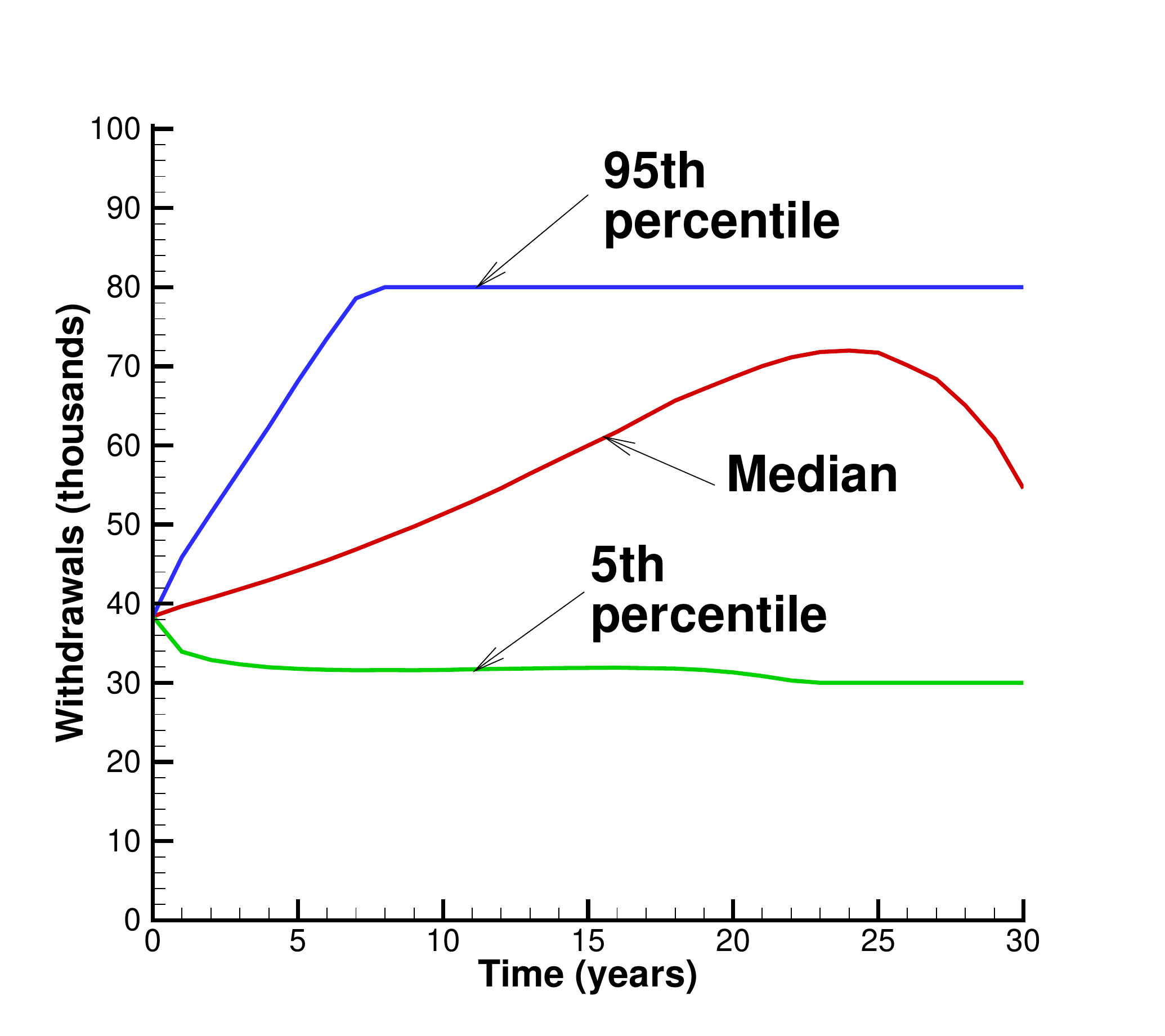}
\caption{Percentiles of withdrawals.}
\label{percentile_q_boot_fig}
\end{subfigure}
\begin{subfigure}[t]{.33\linewidth}
\centering
\includegraphics[width=\linewidth]{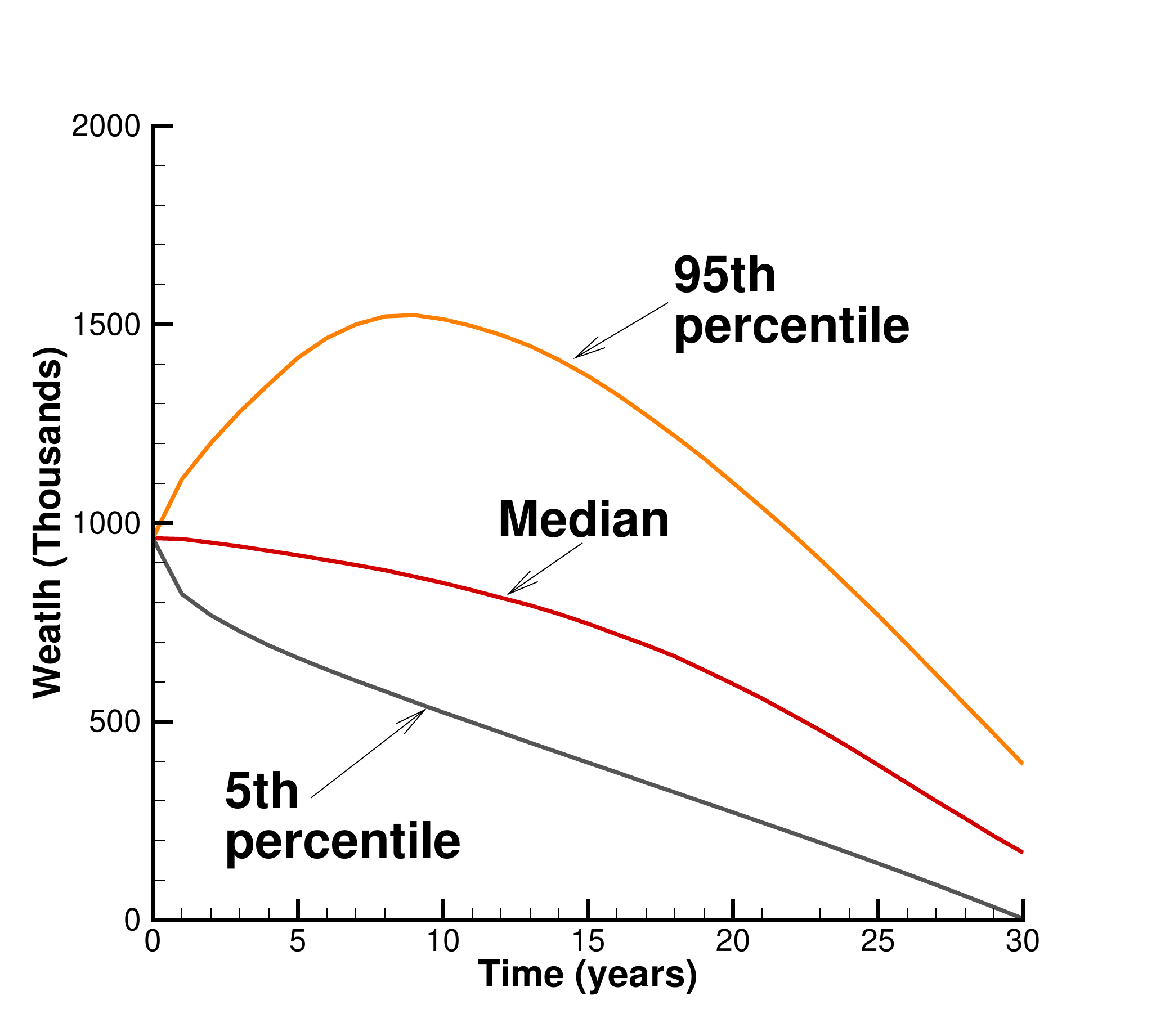}
\caption{Percentiles of wealth.}
\label{percentile_wealth_boot_fig}
\end{subfigure}
}
\caption{Percentiles over time in the historical market of the
fraction invested in the stock index, withdrawals, and wealth for
the scenario from Table~\ref{base_case_1} with ARVA withdrawals
and optimal asset allocation. The scalarization parameter in
equation~(\ref{PCEE_a}) is $\kappa=2.5$ and the stabilization parameter
in equation~(\ref{stable_objective}) is $\epsilon = -10^{-4}$. Based on
$10^5$ bootstrap resamples of the monthly data from 1926:1 to 2018:12.
Units: thousands of dollars.
\label{percentiles_bootstrap_fig}}
\end{figure}

Figure~\ref{percentiles_bootstrap_fig} shows the percentiles of the
optimal controls, withdrawals and wealth throughout the
decumulation period in the historical market with 
$\hat{b} = 2$ years. Figure~\ref{percentiles_bootstrap_fig} is very
similar to the corresponding Figure~\ref{percentiles_synthetic}
for the synthetic market. The median fraction invested
in the stock index increases a little more sharply in
Figure~\ref{percentiles_bootstrap_fig}, and the 5th percentile of this
fraction reaches zero a little later, but these are almost the only
discernible differences. Overall, the close correspondence between
the various panels of these two figures suggests that the parametric
model used when solving for the optimal control is fairly robust as the
historical market makes no assumptions about the processes followed
by the stock and bond indexes.\footnote{However, this is not always true. In
this case, ES (see Table~\ref{optimal_wt_ARVA_30_day_boot_kappa}
with $\hat{b}=2$ years) is about $-41$. As we will see below,
if we try to increase ES to higher values than this, then the controls 
do not appear to be robust.}

We now compare in the historical market the same three strategies
that were considered previously in the synthetic market of
Section~\ref{sec:synthetic_results}, i.e.\ constant withdrawals of
$q=40$ with constant asset allocation weights, ARVA withdrawals with
constant asset allocation weights, and ARVA withdrawals with optimal
asset allocation. Appendix~\ref{detailed_historical} provides tables
of results for these strategies in the historical market with
$\hat{b} = 2$ years; here we present plots based on those results.

\begin{figure}[tb]
\centerline{%
\begin{subfigure}[t]{.5\linewidth}
\centering
\includegraphics[width=\linewidth]{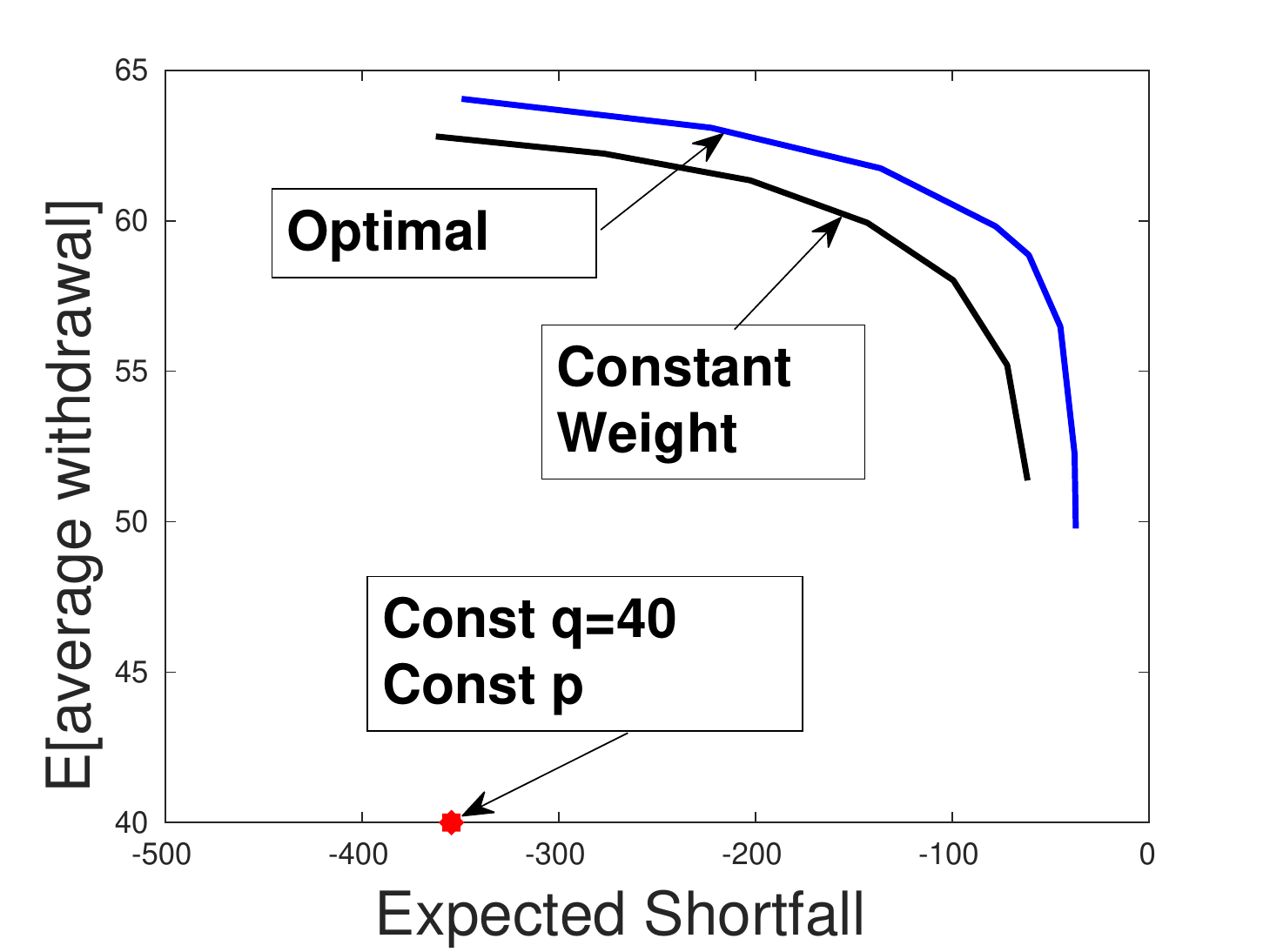}
\caption{ARVA withdrawals with optimal and constant weight
asset allocation, and the single best point for a constant
withdrawal strategy with $q=40$ and constant weight asset
allocation. For this point, $p_{\ell}=0.40$.}
\label{frontiers_bootstrap_fig}
\end{subfigure}
\begin{subfigure}[t]{.5\linewidth}
\centering
\includegraphics[width=\linewidth]{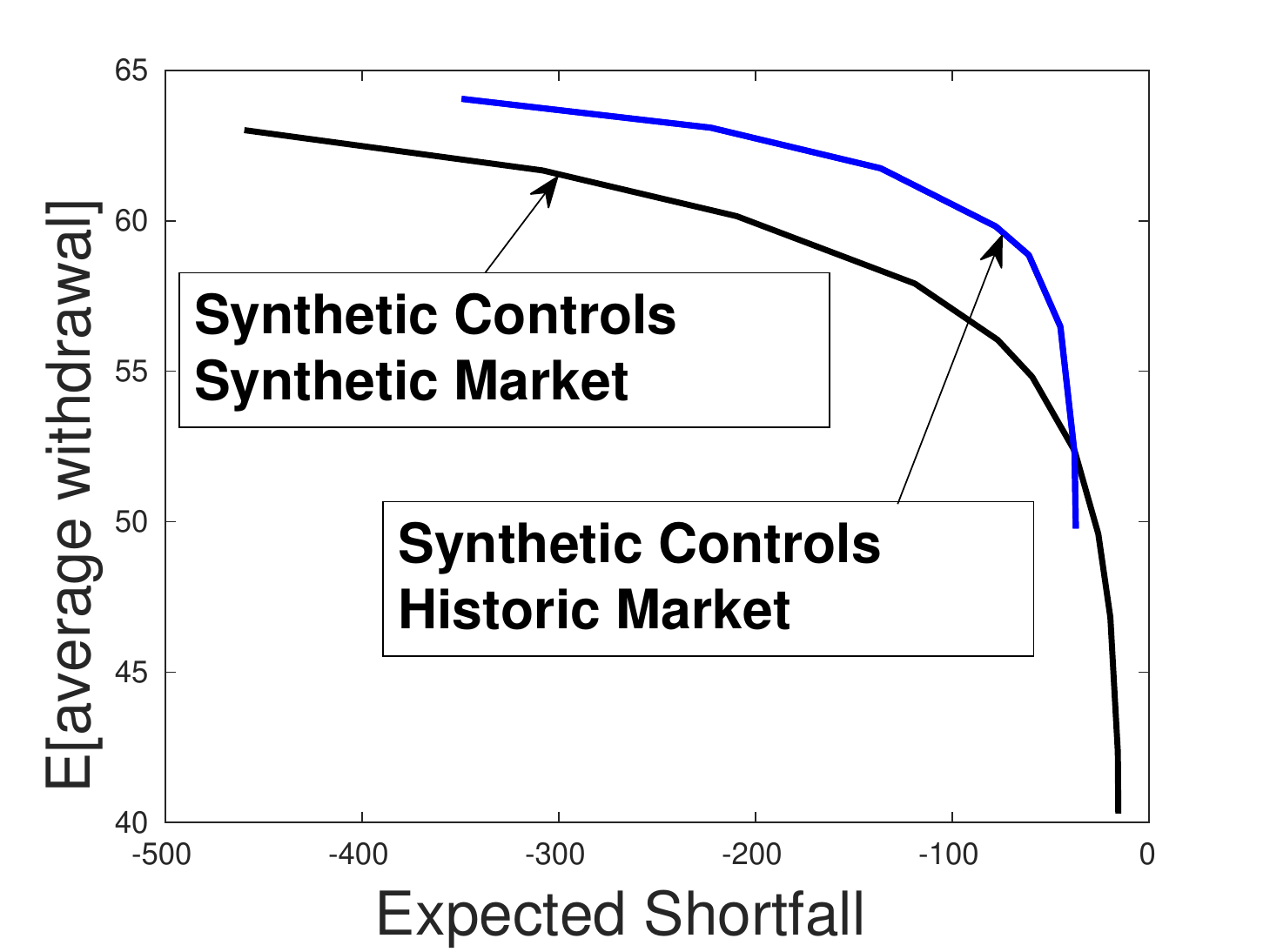}
\caption{ARVA withdrawals with optimal asset allocation, for both
the historical and synthetic markets.}
\label{frontiers_syn_boot_fig}
\end{subfigure}
}
\caption{Efficient frontiers in the historical market for the scenario
from Table~\ref{base_case_1}. All non-Pareto points have been removed.
Units: thousands of dollars.}
\label{frontier_historical_fig}
\end{figure}

The efficient frontiers of expected average withdrawals vs.\ ES in the
historical market are plotted in Figure~\ref{frontiers_bootstrap_fig},
which is analogous to Figure~\ref{compare_frontier_fig} for
the synthetic market. As in Figure~\ref{compare_frontier_fig},
Figure~\ref{frontiers_bootstrap_fig} shows that the ARVA withdrawal
with constant weight asset allocation is a major improvement over
the constant withdrawal with constant asset allocation weights. As
expected, the optimal ARVA withdrawal strategy with optimal asset
allocation continues to plot above the ARVA withdrawal strategy with
constant weight asset allocation, indicating that optimal asset
allocation can provide further significant enhancements. Although
the general picture is the same here in the historical market as
it was in the synthetic market, it is worth pointing out a couple
of specific differences. First, consider the constant withdrawal
strategy with constant asset allocation. In the synthetic market,
the highest ES of about $-284$ for an equity weight of 0.15 (see
Table~\ref{const_wt_const_withdraw_30_day}). This is the best
available point, since withdrawals are constant. In the historical
market, the corresponding ES is about $-355$ for an equity weight of
0.40 (see Table~\ref{const_wt_const_withdraw_30_day_boot}). However,
Figure~\ref{compare_frontier_fig} indicates that in the synthetic market
an ES of $-200$ can be attained with expected average withdrawals of
about 58 for the constant weight case and about 60 for the optimal asset
allocation case. The corresponding values for the historical market in
Figure~\ref{frontiers_bootstrap_fig} with an ES of $-200$ are a little
higher, about 61 for the constant weight case and around 63 for optimal
asset allocation. These values do not constitute the largest gap between
these two frontiers, but they do indicate that ARVA withdrawals (with
either constant weight or optimal asset allocation) perform a bit better
in the historical market relative to the synthetic market, at least for
this level of ES. On the other hand, the performance of the constant
withdrawal strategy is notably worse in the historical market.

%\begin{figure}[b!]
%\begin{center}
%\includegraphics[width=3in]{frontiers_bootstrap}
%\caption{Efficient frontiers in the historical market for ARVA
%withdrawals with optimal asset allocation and ARVA withdrawals
%with constant weight asset allocation for the scenario given in
%Table~\ref{base_case_1}. Also shown is the single best point for a
%constant withdrawal strategy having $q=40$ with constant weight asset
%allocation (for this point, $p_{\ell}=0.40$). Based on $10^5$ bootstrap
%resamples of the monthly data from 1926:1 to 2018:12 with expected
%blocksize $\hat{b}=2$ years. All non-Pareto points have been removed.
%Units: thousands of dollars.
%\label{frontiers_bootstrap_fig}}
%\end{center}
%\end{figure}

A more direct comparison between the synthetic and historical markets
is given in Figure~\ref{frontiers_syn_boot_fig} which plots the
efficient frontiers of expected average withdrawals vs.\ ES for ARVA
withdrawals with optimal asset allocation in both markets, with the
optimal controls having of course been determined in the synthetic
market. The frontier for the historical market plots above
the frontier for the synthetic market if $\text{ES} < -40$. However, the
situation is reversed for $\text{ES} > -40$. This suggests that it is
unreliable to try to achieve very low ES risk in the actual market.
This is not unreasonable, since in order to obtain ES values close to
zero the optimal strategy will depend greatly on the stochastic
market structure. Consequently, it appears that the synthetic market
controls are not robust to parameter uncertainty for $\text{ES} > -40$,
although the controls do appear to be robust otherwise.

%\begin{figure}[tb]
%\begin{center}
%\includegraphics[width=3in]{frontiers_syn_boot}
%\caption{Efficient frontiers in both the synthetic and historical
%markets for ARVA withdrawals with optimal asset allocation for the
%scenario given in Table~\ref{base_case_1}. The optimal controls are
%computed using the synthetic market in each case, stored, and then
%applied to simulated paths in the two markets. The synthetic market
%frontier is based on $2.56 \times 10^6$ Monte Carlo simulated paths. The
%historical market frontier is based on $10^5$ bootstrap resamples of the
%monthly data from 1926:1 to 2018:12 with expected blocksize $\hat{b} =
%2$ years. All non-Pareto points have been removed. Units: thousands of
%dollars.
%\label{frontiers_syn_boot_fig}}
%\end{center}
%\end{figure}

\section{Conclusions}
For both parametric model simulations and bootstrap resampling of
the historical data, the ARVA withdrawal strategy with constant
asset weights and minimum/maximum withdrawal constraints outperforms
a constant withdrawal strategy with constant asset weights based
on expected average withdrawals and expected shortfall criteria.
This is consistent with results from the practitioner literature
\citep[e.g.][]{Pfau_2015} which show that withdrawal variability
can significantly improve performance in cases with constant weight
asset allocation. However, we also show that the ARVA withdrawal
strategy can be further improved by dynamically choosing the equity
weight. This strategy is determined by maximizing an expected total
withdrawals/expected shortfall objective function using dynamic
programming, assuming a parametric model of historical asset returns.
As long as the desired expected shortfall is not unrealistically large,
this strategy is robust to parameter misspecification, as verified by
tests using bootstrapped resampled historical data.

Remarkably, the optimal dynamic ARVA strategy continues to outperform
the constant withdrawal/constant weight strategy, even if the
\emph{minimum} ARVA withdrawal is set equal to the constant withdrawal
in the latter strategy. These results indicate that if an investor
in the decumulation stage of a DC plan is prepared to allow some
variability in withdrawals, significant improvements can be obtained in
both expected total withdrawals and expected shortfall.

\appendix
\section*{Appendix}

\section{Calibration of Model Parameters}
\label{calibration_section}

This appendix discusses the estimation of the parameters of the
jump diffusion processes for the stock and bond indexes given
by equations \eqref{eq:dist_stock}, \eqref{jump_process_stock},
\eqref{jump_process_bond}, and \eqref{eq:dist_bond}. Recall that the
equity index is the CRSP value-weighted stock index while the bond index
is the CRSP 30-day T-bill index, and that both of these indexes are
adjusted for inflation by using the CPI.

Jumps in the data are identified using the
thresholding technique described in \citet{mancini2009} and
\citet{contmancini2011}. Let $\Delta
\hat{X}_i$ be the detrended $\log$ return in period $i$, with period
time interval $\Delta t$. Suppose we have an estimate for the diffusive
volatility component $\hat{\sigma}$. Then we detect a jump in period $i$
if $\left|\Delta \hat{X}_i \right| > \beta \: \hat{\sigma}  \sqrt{\Delta t}$.
We choose $\beta = 3$ in this paper (note that $\Delta t$ is fixed).  For justification
for this parameter selection, see \citep{shimizu2013,Dang2015a,forsyth_robust_2017}.
For details describing the recursive algorithm used to determine $\hat{\sigma}$,
see \citet{forsyth_robust_2017}.

\begin{figure}[tb]
\centerline{%
  \begin{subfigure}[b]{.5\linewidth}
    \centering
    \includegraphics[width=\linewidth]{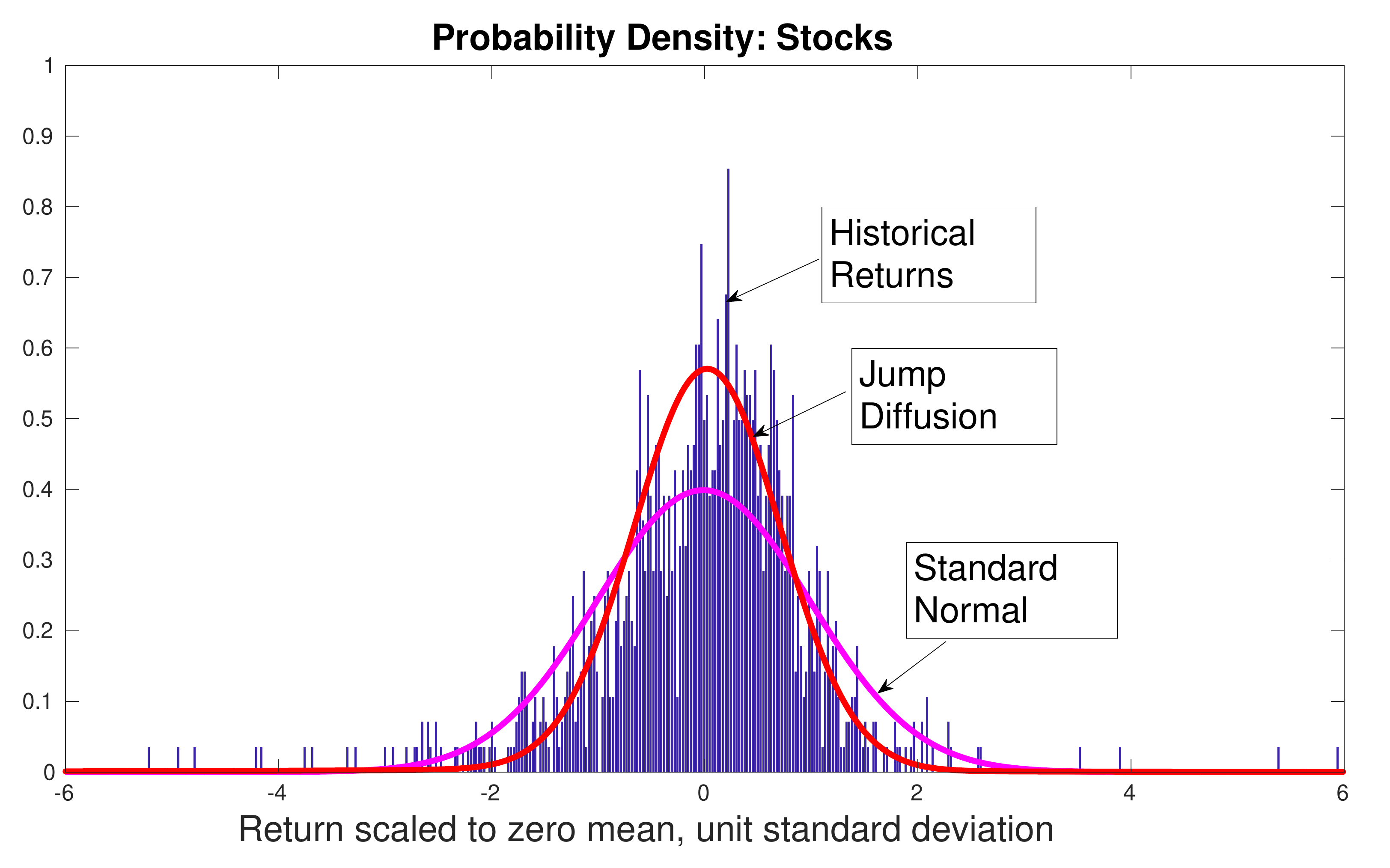}
                % same scale
    \caption{Log returns and densities, stock index.\label{density_1}}
  \end{subfigure}
  \begin{subfigure}[b]{.5\linewidth}
    \centering
    \includegraphics[width=\linewidth]{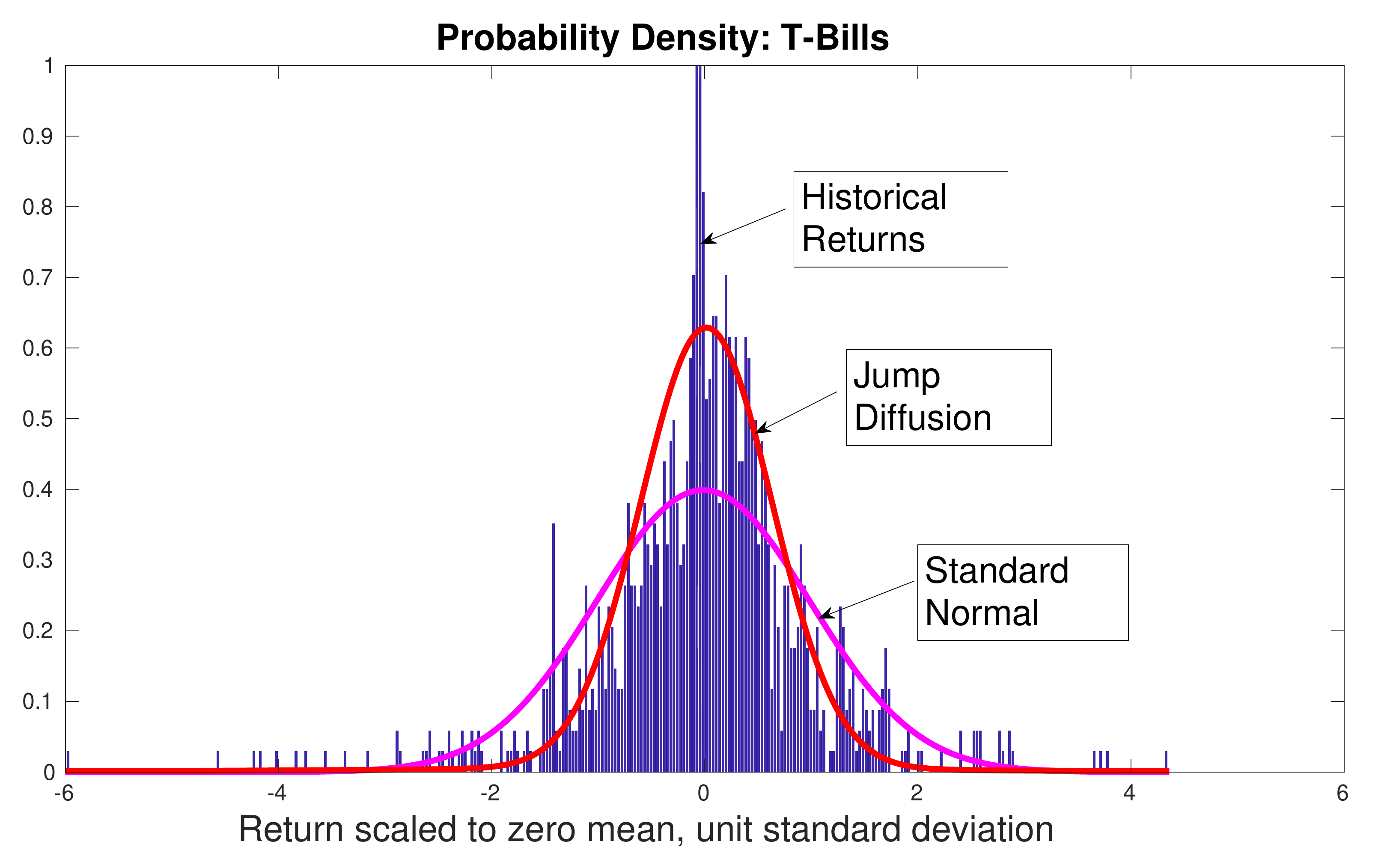}
             %same scale
    \caption{Log returns and densities, T-bill index.\label{density_2}}
  \end{subfigure}
}
\caption{Actual and fitted log returns for the CRSP value-weighted
equity index and 30-day T-bill indexes. Monthly data from
1926:1-2018:12, scaled to zero mean and unit standard deviation. A
standard normal density and the fitted double exponential jump diffusion
density (threshold, $\beta = 3$) are also shown.
\label{monthly_hist_1}}
\end{figure}

Figure~\ref{density_1} shows a histogram of the monthly $\log$ returns
from the value-weighted CRSP stock index, scaled to zero mean and unit
standard deviation. We superimpose a standard normal density onto
this histogram, as well as the fitted density for the double
exponential jump diffusion model. Figure~\ref{density_2} shows the
equivalent plot for the 30-day T-bill index. 

During the sample period of 1926:1-2018:12 (monthly), the filtering
algorithm identified 30 stock index jumps and 48 T-bill index jumps. Of
these cases, just 5 were identified as occurring in the same month for
both stocks and bonds, all in the 1930s. This supports our modelling
assumption of no dependence between the jump intensities or jump
distributions of the two indexes, though we do allow for correlated
Brownian motion terms in the parametric model.

\section{Historical Market: Detailed Results}
\label{detailed_historical}

This appendix presents detailed results for the historical market
bootstrap resampling tests with expected blocksize $\hat{b} = 2$ years.
Table~\ref{const_wt_const_withdraw_30_day_boot} shows the results for a
constant withdrawal ($q=40$) strategy with constant equity weight asset
allocation, analogous to Table~\ref{const_wt_const_withdraw_30_day}
in the synthetic market. Table~\ref{const_wt_ARVA_30_day_boot} gives
results for ARVA withdrawals with constant equity weight asset
allocation, analogous to Table~\ref{const_wt_ARVA_30_day} in the
synthetic market. Finally, Table~\ref{optimal_wt_ARVA_30_day_boot}
presents results in the historical market for ARVA withdrawals and
optimal asset allocation (the optimal control is computed by solving the
pre-commitment EW-ES problem~(\ref{PCEE_a}) in the synthetic market).
This table is analogous to Table~\ref{optimal_wt_ARVA_30_day} for the
synthetic market.

\begin{table}[htb!]
\begin{center}
\begin{tabular}{cD{.}{.}{2}D{.}{.}{2}} \toprule
Equity Weight $p_{\ell}$ & \multicolumn{1}{c}{ES ($\alpha=5\%$)} &
\multicolumn{1}{c}{Median$[W_T]$} \\ \midrule
0.0 & -550.33 & -191.87 \\
0.1 & -461.16 & -52.68 \\
0.2 & -394.73 & 113.56 \\
0.3 & -358.56 & 317.35 \\
0.4 & -354.67 & 562.04 \\
0.5 & -378.58 & 850.23 \\
0.6 & -425.71 & 1177.31 \\
0.7 & -490.42 & 1548.45 \\
0.8 & -568.29 & 1956.86 \\
0.9 & -655.39 & 2381.87 \\
1.0 & -750.09 & 2823.11 \\ \bottomrule 
\end{tabular}
\caption{Historical market results for constant withdrawals
with constant weights, i.e.\ assuming the scenario given in
Table~\ref{base_case_1} except that $q_{\max} = q_{\min} = 40$, and
$p_{\ell} = \text{\emph{constant}}$ in equation~(\ref{PCEE_b}). Units:
thousands of dollars. Statistics based on $10^5$ bootstrap resamples of
the monthly data from 1926:1 to 2018:12 with expected blocksize $\hat{b}
= 2$ years.
\label{const_wt_const_withdraw_30_day_boot}}
\end{center}
\end{table}

\begin{table}[htb]
\begin{center}
\begin{tabular}{cD{.}{.}{2}D{.}{.}{2}D{.}{.}{2}} \toprule
Equity Weight $p_{\ell}$ & \multicolumn{1}{c}{ES ($\alpha=5\%$)} &
  \multicolumn{1}{c}{$\text{EW}/(M+1)$} &
  \multicolumn{1}{c}{Median[$W_T$]} \\ \midrule
0.0 & -227.41 & 35.79 & -13.79 \\
0.1 & -151.74 & 38.53 & 31.44 \\
0.2 & -98.37  & 42.27 & 64.71 \\
0.3 & -69.44  & 46.79 & 90.45 \\
0.4 & -61.86  & 51.37 & 111.55 \\
0.5 & -72.20  & 55.20 & 137.97 \\
0.6 & -99.58  & 58.02 & 170.37 \\
0.7 & -143.23 & 59.93 & 269.27 \\
0.8 & -202.74 & 61.34 & 493.52 \\
0.9 & -277.09 & 62.23 & 766.16 \\
1.0 & -362.60 & 62.80 & 1069.33 \\ \bottomrule
\end{tabular}
\caption{Historical market results for ARVA withdrawals
with constant weights, i.e.\ assuming the scenario given in
Table~\ref{base_case_1} except that $p_{\ell} = \text{\emph{constant}}$
in equation~(\ref{PCEE_b}). There are $M=30$ rebalancing dates and $M+1$
withdrawals. Units: thousands of dollars. Statistics based on $10^5$
bootstrap resamples of the monthly data from 1926:1 to 2018:12 with
expected blocksize $\hat{b} = 2$ years.
\label{const_wt_ARVA_30_day_boot}}
\end{center}
\end{table}

\begin{table}[htb]
\begin{center}
\begin{tabular}{rD{.}{.}{2}D{.}{.}{2}D{.}{.}{2}D{.}{.}{3}} \toprule
\multicolumn{1}{c}{$\kappa$} & \multicolumn{1}{c}{ES ($\alpha=5\%$)} &
  \multicolumn{1}{c}{$\text{EW}/(M+1)$} &
  \multicolumn{1}{c}{Median[$W_T$]} &
  \multicolumn{1}{c}{$\sum_i \text{Median}(p_i)/M$} \\ \midrule
0.1   & -349.50 & 64.05 & 258.80 & .466 \\
0.25  & -222.76 & 63.09 & 253.57 & .473 \\
0.4   & -136.43 & 61.74 & 247.42 & .482 \\
0.7   & -78.02  & 59.81 & 239.01 & .464 \\
1.0   & -61.23  & 58.86 & 230.46 & .452 \\
1.75  & -45.17  & 56.48 & 204.19 & .432 \\
2.5   & -40.80  & 55.15 & 180.32 & .416 \\
5.0   & -37.96  & 52.26 & 135.64 & .382 \\
10.0  & -37.34  & 49.77 & 101.99 & .335 \\
100.0 & -42.87  & 43.22 & 53.70  & .214 \\ \bottomrule
\end{tabular}

\caption{Historical market results for ARVA withdrawals with optimal
asset allocation based on the scenario given in Table~\ref{base_case_1}
for various values of $\kappa$. The optimal control that solves the
pre-commitment EW-ES problem~(\ref{PCEE_a}) is computed in the synthetic
market using the algorithm given in Section~\ref{algo_section}, stored,
and then applied to bootstrap resamples of the historical data. There
are $M=30$ rebalancing dates and $M+1$ withdrawals. Units: thousands of
dollars. Statistics based on $10^5$ bootstrap resamples of the monthly
data from 1926:1 to 2018:12 with expected blocksize $\hat{b}=2$ years.
The stabilization parameter in equation~(\ref{stable_objective}) is
$\epsilon = -10^{-4}$.
\label{optimal_wt_ARVA_30_day_boot}}
\end{center}
\end{table}

\newpage

\begin{singlespace}
\bibliographystyle{chicago}
\bibliography{paper}
\end{singlespace}

\end{document}